\documentclass[showpacs,preprintnumbers,amsmath,amssymb,APSl,prd,nofootinbib,superscriptaddress, twocolumn]{revtex4-2}

\usepackage{dcolumn}
\usepackage{bm}
\usepackage{ifpdf}
\usepackage{hyperref}
\usepackage{bm}
\usepackage{xcolor,color,graphicx,graphics}
\usepackage[spanish,english]{babel}
\usepackage[latin1]{inputenc}
\usepackage[OT1]{fontenc}
\usepackage{latexsym,amssymb,amsmath,amsfonts}
\usepackage{makeidx}
\usepackage{epsfig,subfigure}
\usepackage{natbib}
\usepackage{epstopdf}
\usepackage{mathrsfs}
\hypersetup{colorlinks=true, linkcolor=blue, citecolor=green}
\usepackage{enumerate}
\usepackage{xcolor}
 \usepackage{multirow}

\definecolor{red}{rgb}{1,0,0}

\def\+{^\dagger}

\def\<{\leftarrow}
\def\>{\rightarrow}

\def\({\left(}
\def\){\right)}

\newcommand{\bi}{\begin{itemize}} 				\newcommand{\ei}{\end{itemize}}
\newcommand{\benu}{\begin{enumerate}} 		\newcommand{\enu}{\end{enumerate}}
\newcommand{\bd}{\begin{dinglist}{0}}     \newcommand{\ed}{\end{dinglist}}
\newcommand{\bfig}{\begin{figure}[htbp]}  \newcommand{\efig}{\end{figure}}
        			
\newcommand{\bc}{\begin{center}} 				  \newcommand{\ec}{\end{center}}
\newcommand{\be}{\begin{equation}} 				\newcommand{\ee}{\end{equation}}
\newcommand{\bsub}{\begin{subequations}}  \newcommand{\esub}{\end{subequations}}
\newcommand{\ben}{\begin{eqnarray}} 			\newcommand{\een}{\end{eqnarray}}
\newcommand{\ba}[1]{\begin{array}{#1}} 		\newcommand{\ea}{\end{array}}
\newcommand{\bea}{\begin{equation}\begin{array}{rcl}}
\newcommand{\eea}{\end{array}\end{equation}}

\begin{document}
\title{Multi-ring images of thin accretion disk of a regular naked compact object}

\author{Merce Guerrero} \email{merguerr@ucm.es}
\affiliation{Departamento de F\'isica Te\'orica and IPARCOS,
	Universidad Complutense de Madrid, E-28040 Madrid, Spain}
	\author{Gonzalo J. Olmo} \email{gonzalo.olmo@uv.es}
\affiliation{Departamento de F\'{i}sica Te\'{o}rica and IFIC, Centro Mixto Universidad de Valencia - CSIC.
Universidad de Valencia, Burjassot-46100, Valencia, Spain}
\affiliation{Departamento de F\'isica, Universidade Federal do Cear\'{a} (UFC), Campus do Pici, Fortaleza - CE, C.P. 6030, 60455-760, Brazil}
\author{Diego Rubiera-Garcia} \email{drubiera@ucm.es}
\affiliation{Departamento de F\'isica Te\'orica and IPARCOS,
	Universidad Complutense de Madrid, E-28040 Madrid, Spain}
	\author{Diego S\'aez-Chill\'on G\'omez} \email{diego.saez@uva.es}
\affiliation{Department of Theoretical Physics, Atomic and Optics, Campus Miguel Delibes, \\
University of Valladolid UVA, Paseo Bel\'en, 7, 47011 - Valladolid, Spain}

\date{\today}
\begin{abstract}
We discuss the importance of multi-ring images in the optical appearance of a horizonless spherically symmetric compact object, when illuminated by an optically thin accretion disk. Such an object corresponds to a sub-case of an analytically tractable extension of the Kerr solution dubbed as the {\it eye of the storm} by Simpson and Visser in [JCAP \textbf{03} (2022) 011], which merits in removing curvature singularities via an asymptotically Minkowski core, while harbouring both a critical curve and an infinite potential barrier at the center for null geodesics. This multi-ring structure is induced by light rays winding several times around the object, and whose luminosity is significantly boosted as compared to the Schwarzschild solution by the modified shape of the potential. Using three toy-profiles for the emission of an infinitely thin disk, truncated at its inner edge (taking its maximum value there) and having different decays with the distance, we discuss the image created by up to eight rings superimposed on top of the direct emission of the disk as its edge is moved closer to the center of the object. Our results point out to the existence of multi-ring images with a non-negligible luminosity in shadow observations when one allows for the existence of other compact objects in the cosmic zoo beyond the Schwarzschild solution. Such multi-ring images could be detectable within the future projects on very-long baseline interferometry.

\end{abstract}

\maketitle

\section{Introduction}
\label{intro}

The detection by the Event Horizon Telescope in 2017  of the image created by the accreting plasma around the central supermassive object of the M87 galaxy \cite{EventHorizonTelescope:2019dse} and, very recently, also that of the central source of our own Milky Way -  Sgr A* - \cite{EHT2022PaperI}, has tremendously enhanced our chances to verifying the Kerr paradigm or, alternatively, to search for the potential existence of other ultra-compact objects in the cosmic zoo. In this sense, while the background geometry is the solely responsible for the  {\it critical curve}, i.e. the locus of unstable null geodesics intersecting the observer's screen \cite{Cunha:2018acu,Gralla:2019drh}, in astrophysically realistic scenarios  the optical appearance of every compact object is highly dependent on the optical and geometrical properties of the emission source (i.e. its accretion disk) illuminating it. Disentangling the contributions from each other in such an image is one of the main challenges in the field, since it harbours important clues to test the nature of compact objects and, consequently, of the Kerr-paradigm, which in turn would help to reveal the nature of the gravitational interaction on its strong-field regime \cite{Ohgami:2015nra,Olivares:2018abq,Shaikh:2019hbm,Held:2019xde,Saurabh:2020zqg,Peng:2020wun,Zeng:2020vsj,Li:2021ypw,Li:2021riw,Zeng:2021mok,Guo:2021wid,Kocherlakota:2022jnz}.

According to General Relativity (GR) predictions, both in terms of analytical models \cite{Perlick:2021aok} and as outcomes of magnetohydrodynamics simulations \cite{Gold:2020iql}, the image of every such object when the main source of illumination is its accretion disk is strongly dominated by a broad ring of radiation originated by the direct emission of the disk, namely, light paths turning at most half an orbit around the object, and which bounds a central brightness depression (the shadow). Measurements of the diameter of this bright ring can act, under certain circumstances and after proper calibration, as an indication of the size of the shadow itself, thus allowing to rule out some candidates whose predictions are in open contradiction with this result \cite{EHT2022PaperVI}, but by itself it does not allow to conclusively determine the nature of the shadow caster \cite{Wielgus:2021peu}. This is so because the direct emission basically conveys the image of the source \cite{Gralla:2020pra} and, therefore, two compact objects can cast very similar (direct) images  \cite{Lima:2021las,Herdeiro:2021lwl}. Nonetheless, shadow observations may allow to constraint black hole charges \cite{EventHorizonTelescope:2021dqv}, the parametrized deviations with respect to the Kerr/Schwarzschild solution \cite{Volkel:2020xlc} as well as a bunch of well motivated deviations with respect to such a solution \cite{Vagnozzi:2022moj}.

When the disk is optically thin, that is, transparent to its own radiation,  photons turning more than one-half orbit around the compact object produce a thin photon ring besides the direct emission. If the disk is spherically symmetric, such photon ring converges to the critical curve itself and delimits the outer edge of the shadow  \cite{Narayan:2019imo}, while if the disk is geometrically thin (or even thick, see the recent work \cite{Vincent:2022fwj}), then it is broken instead into an infinite sequence of self-similar photon rings which have performed several (half-) loops around the critical curve, and which are stacked on the direct emission \cite{Gralla:2019xty}. Indeed, in the latter case, if the inner edge of the disk penetrates deep enough inside the critical curve, then the minimum size of the shadow is not necessarily tied to the one of the critical curve but can be strongly reduced instead, which in a typical black hole case with a disk extending up to its event horizon is dubbed as the inner shadow \cite{Chael:2021rjo}. As for photon rings, their size and location can be quite different depending on the features of the shadow caster, while their corresponding luminosity is also tightly attached to the interaction between the background geometry and the astrophysics of the disk. However, in the Kerr/Schwarzschild black hole case, higher-order rings beyond the third half-orbit are typically neglected in the optical appearance on the basis that their contributions to the total luminosity  turn out to be exponentially suppressed as compared to that of the direct emission \cite{Bisnovatyi-Kogan:2022ujt}. Moreover, mild deviations from such a solution are expected not to modify significantly this conclusion. Nonetheless, since the cosmic zoo of compact objects could be populated by exotic beasts (e.g. boson/Proca stars, gravastars, traversable wormholes, etc \cite{Cardoso:2019rvt}) exhibiting both qualitatively and quantitatively large departures from the Schwarzschild expectations (at least in certain regimes), one should pay careful attention to the possibility of observing multi-ring images with unexpected features, as they may offer prospects of new physics hidden in shadow observations  \cite{Ayzenberg:2022twz}.

Modifications with respect to the Kerr/Schwarzschild solutions have been sought for according to two main avenues: i) parameterized, model-agnostic, deviations with respect to such a solution, whose parameters can be constrained according to different arrays of observations \cite{Param1,Param2}, ii) exact or numerical solutions out of well defined theories of the gravitational and/or matter fields \cite{Eichhorn:2021iwq,Daas:2022iid}. For the sake of their shadow images, discerning the nature of the background geometry via the disk's direct emission is subject to some limitations due to the still poorly understood physics of the plasma \cite{Lara}. This makes the multi-ring structure, whose location and size are more dependent on the geometry of the space-time and less on the accretion disk physics (an effect more pronounced as more half-orbits are considered, i.e., as light rays approach closer to the critical curve), to play a more prominent role in testing such a geometry in a less astrophysical model-dependent way \cite{Hou:2022gge}. Therefore, a characterization of such a multi-ring structure in different geometries represents a promising avenue to test the strong-field regime of the gravitational interaction \cite{Gralla:2020srx,Carballo-Rubio:2022bgh}.

The main aim of this work is to explore the multi-ring structure of a toy-model geometry  to argue that the assumption on the faintness of the higher-order photon rings formed by illumination from an  optically and geometrically thin accretion disk  can be substantially modified when the effective potential for geodesics has a qualitatively different shape near the center of the compact object as compared to Schwarzschild expectations. Specifically we are considering the family of geometries (in the spherically symmetric limit) originally proposed by Ghosh in \cite{Ghosh:2014pba} (see also \cite{Culetu:2014lca}) and further developed by  Simpson and Visser in \cite{Simpson:2021dyo,Simpson:2021zfl} under the nickname of {\it eye of the storm}. While it describes a tractable extension of the Kerr space-time, for the sake of this paper its main novelty lies in the fact that the effective potential for null geodesics has both a local maximum (a critical curve) and a local minimum (an anti-photon sphere), while displaying an infinite potential barrier at its center. While the overcharged Reissner-Nordstr\"om solution (when $M^2<Q^2<(9/8)M^2$) displays the same qualitative behaviour, it bears the burden of harbouring an unpleasant naked singularity at is center\footnote{Other naked singularities show similar features in their images as those of the regular model studied here, as in the case Janis-Newman-Winicour  \cite{Gyulchev:2020cvo} and Gauss-Bonnet \cite{Gyulchev:2021dvt} space-times.}, while such a feature is absent in the present Simpson-Visser (SV) model,  being replaced by an asymptotically Minkowski-like core. This follows the trail of the  so-called {\it de Sitter cores} \cite{Ansoldi:2008jw}, a  mechanism frequently invoked in the literature to regularize curvature singularities. On the one hand, while the idea of the presence of a local maximum and minimum in the effective potential is not new, since it can be implemented under several mechanisms \cite{Wielgus:2020uqz,Wang:2020emr,Guerrero:2021pxt,Peng:2021osd,Shaikh:2019jfr,Shaikh:2019itn,Junior:2021dyw,Rodrigues:2022mdm,Guo:2022muy}, the reflective potential barrier of this object (which resembles the Type II-solutions recently considered in \cite{Daas:2022iid}) introduces a significant boost in the luminosity of its multi-ring structure, making several them to be visible even at naked eye. While this model is arguably not a viable alternative to represent the shadow caster of the EHT observations, it serves as a {\it proof of concept} for the higher-order ring images to act as observational discriminators with respect to canonical GR predictions.

The paper is organized as follows: Sec. \ref{model} introduces the spacetime metric that plays the central role of the paper and the basic geodesic equations, while the main analysis and results for the multi-ring images are gathered in Sec. \ref{HOrings}. Finally, Sec. \ref{conclusions} is devoted to the conclusions and open questions of the manuscript.

\section{The model}
\label{model}

We consider a static, spherically symmetric geometry of the form
\begin{equation} \label{eq:motion}
ds^2=-A(r)dt^2+A^{-1}(r)dr^2+r^2d\Omega^2 \ ,
\end{equation}
where $d\Omega^2=d\theta^2 +\sin^2 \theta d\varphi^2$ is the usual angular element in the unit sphere, while the single free metric function $A(r)$ is given by the non-rotating limit of the SV family of configurations  \cite{Simpson:2021dyo}  as
\begin{equation} \label{eq:metric}
A(r)=1-\frac{2Me^{-l/r}}{r} \ ,
\end{equation}
where $l >0$ is a new scale parameterizing the deviations with respect to the Schwarzschild solution, to which these configurations reduce in the asymptotic limit, $r \to \infty$. Such deviations do manifest as we approach the center of the solutions, where  $A(r \to 0) \to 1$, which in turn removes the point-like central curvature singularity, replacing it by a Minkowski-like core. This model can be taken in a  theory-agnostic approach as a proxy for the features induced by the new behaviour of the metric function\footnote{From the inferred size of the shadow's size of Sgr A* by the EHT Collaboration \cite{EHT2022PaperVI}, the authors of  \cite{Vagnozzi:2022moj} argue on the compatibility of this kind of models with such observations.}. Horizons are present in this model at the locations $r_h$ as far as the following equation owns positive and real roots:
\begin{equation}
r_h\text{e}^{l/r_h}=2M\ ,
\end{equation}
which holds provided that $l/M \lesssim 0.73576$, corresponding to black holes without curvature singularities. Otherwise in absence of horizons the metric (\ref{eq:motion}) leads to regular naked compact objects.

\begin{figure}[t!]
\includegraphics[width=8.0cm,height=5.5cm]{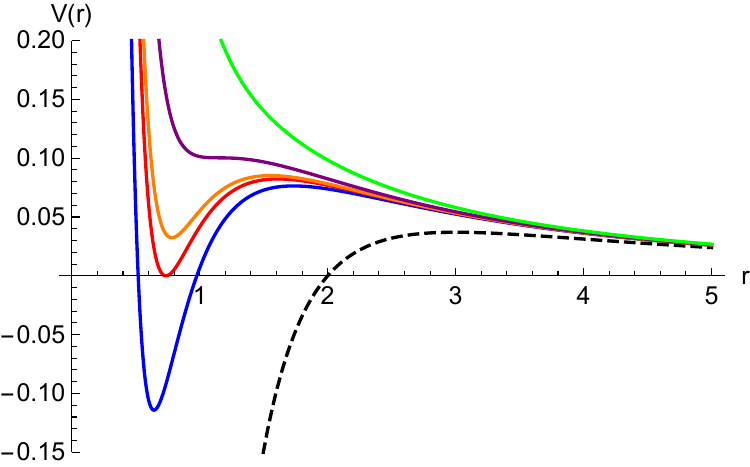}
\caption{The effective potential (in units of $M=1$) for $l=0$ (Schwarzschild, dashed black), $l=0.7$ (blue), $l \approx 0.73576$ (red), $l=0.75$ (orange), $l=0.8$ (purple) and $l=1$ (green).}
\label{fig:potential}
\end{figure}

The motion of photons is assumed to follow null geodesics of the background metric given by the line element  (\ref{eq:motion}), being thus governed by the equation (here a dot represents a derivative with respect to the re-scaled affine parameter $\lambda$)
\begin{equation} \label{eq:geoeq}
\dot{r}^2=\frac{1}{b^2}-V_{eff}(r) \ ,
\end{equation}
where $b=L/E$ is the impact parameter defined by the ratio of the photon's conserved  energy, $E=-A\dot{t}$, and angular momentum (we take $\theta=\pi/2$, a freedom granted by the spherical symmetry of the system, without any loss of generality), $L=r^2 \dot{\phi}$, and the effective potential is given by $V_{eff}(r)=A(r)/r^2$. From (\ref{eq:metric}) one can see that this potential strongly departs from the Schwarzschild one at the center of the solutions, where  $V(r\to 0) \to +\infty$ (instead of going to $- \infty$). Moreover, a local maximum and minimum is present  provided that $l <0.8$ (see Fig. \ref{fig:potential}). Combining the parameter space for horizon and critical curves, one finds a particularly interesting region within the range  $ 0.73576 \lesssim l/M < 0.8$ in which we have a kind of regular naked object with an accessible critical curve as well as an anti-photon sphere\footnote{While the presence of anti-photon spheres can be able to trigger a non-perturbative instability \cite{Keir:2014oka,Cardoso:2014sna,Cunha:2017qtt,Guo:2021bcw,Guo:2021enm}, the time scales at which this may occur are model-dependent, thus not being a completely unsurmountable argument against the viability of solutions holding them.}, supplemented with a infinite reflective potential barrier at its center. This is a sufficiently new structure from a qualitative point of view to motivate the investigation of its cast images and from now on we shall dismiss every other configuration of this family. For the sake of our computations, we shall take $l=3/4$ (using $M=1$, which we shall assume in the rest of the paper).

According to Eq.(\ref{eq:geoeq}), every photon travelling from asymptotic infinity will have a turning point at a distance $r_0$ from the center of the object, given by the vanishing of the right-hand side of that equation, i.e., $1/b^2=V_{eff}(r_0)$. The minimum radius $r_m$ such a photon can approach to is given by the location of the (local) maximum of the potential, being dubbed as the {\it photon sphere}, and which corresponds, in the observer's screen, to the {\it critical curve} given by the impact parameter 
\begin{equation}
b_c=\frac{r_{m}}{\sqrt{A(r_{m})}} \ .
\end{equation}
Since this is the locus of unstable bound orbits, it effectively splits the space of light rays issued from the observer's screen into two classes: those with $b>b_c$ are deflected at  $r_0$ back towards asymptotic infinity, while those with $b<b_c$ will inspiral down towards the center of the object (thus meeting the event horizon in a black hole case). For the model and parameter's choice considered in this work, the critical curve/photon sphere are characterized by $b_c \approx 3.4263$ and $r_m \approx 1.5529$, which are significantly smaller than their corresponding values in the Schwarzschild case ($b_c=3\sqrt{3}\approx 5.1962$ and $r_m=3$). Despite the fact that these values of the critical curve make this solution hardly compatible with current EHT observations, the novel optical appearance due to the higher-order rings distribution of this object, which can act as a proxy of the expected features of more consistent theoretical objects having similar properties, is of enough importance to deserve its analysis.

\section{Multi-ring images}
\label{HOrings}

\subsection{Ray-tracing} \label{sec:raytrac}

We now implement a ray-tracing procedure in order to characterize the different trajectories contributing to the image of the object on the observer's screen. For this purpose, we first suitably rewrite (\ref{eq:geoeq}) as
\begin{equation} \label{eq:ray-trac}
\frac{d\phi}{dr}=\mp \frac{b}{r^2\sqrt{1-\frac{b^2A(r)}{r^2}}} \ .
\end{equation}
In the ray-tracing procedure one is only interested on those light rays reaching the observer's screen, so that Eq.(\ref{eq:ray-trac}) is integrated backwards from there and the corresponding trajectories are indexed by the number of (half-) orbits, $n\equiv \tfrac{\phi}{2\pi}$ (following the notation of \cite{Gralla:2019xty}). Furthermore we conveniently orient our system so that the observer is located in the North pole and the equatorial plane lies in the $y$ axis. In this language, a light ray travelling from left to right on this setup without any lensing has $n=1/2$ orbits. We shall afterwards assume an infinitely thin accretion disk living precisely on the equatorial plane and providing the main source of illumination of the SV object  (so that the image seen from the observer will be face-on). Moreover, since the disk will also be assumed to be optically thin, every light ray performing a number of $n$ half-orbits around the object will be able to contribute to the final image of it.

Under the conditions above, the multi-ring structure created by the illuminating disk will be indexed by an integer number $m$ that counts the number of intersections of a particular light ray with it, i.e.:
\begin{equation}
\frac{m}{4}-\frac{1}{4} \leq n < \frac{m}{2}+\frac{1}{4} \ ,
\end{equation}
except for the first case $m=1$, for which the lower limit corresponds to $n=1/2$ (i.e., in terms of half orbits). In the usual Schwarzschild solution, every such trajectory will be in a one-to-one correspondence with the luminous rings appearing in the observer's screen, therefore both usually receiving the name of {\it photon rings}. In our case, as we shall see, this correspondence is broken, and large enough values of $m$ will have associated more than one ring due to presence of the potential well depicted in Fig. \ref{fig:potential}. Given the discrete character of $m$, we shall dub it as the emission {\it mode} number of the trajectory, and reserve the word {\it photon ring} for the luminous rings appearing in the observer's screen, The bottom line of this discussion is that we can use this number in order to classify the different contributions to the optical appearance\footnote{This notation is slightly different from the (perhaps) more canonical in the community, in which $m=0$ is reserved for the direct emission and $m=1,2,\ldots$ for photon ring images, see e.g. \cite{Vincent:2022fwj}. Given the degeneracy between $m$ numbers and photon ring images in our case, we find it clearer to use $m$ for the number of intersections with the disk, which means that our $m$ is always one unit greater than the usual convention.}:
\begin{itemize}
\item Direct emission: it represents light rays that intersect the accretion disk $m=1$ times ($1/2 \leq n < 3/4$) and essentially reproduces the source shape. It thus describes the accretion disk features rather than those of the background geometry and its photon sphere.
\item Lower-order emission modes: they correspond to $m=2, 3$, which are those created by light rays which have performed $3/4 \leq n<5/4$ and $5/4 \leq n<7/4$ loops, i.e., light rays coming from the back side of the disk and from the front after turning a whole loop around the object, respectively. In the Schwarzschild geometry these  lower-order trajectories contribute to exponentially diminished images approaching the critical curve \cite{Bisnovatyi-Kogan:2022ujt} in such a way that beyond $n \geq 7/4$ all further images contribute negligibly to the luminosity of the object, so that they all can be accumulated in the mode $m=3$.
\item Higher-order emission modes: $m>3$. They are triggered by the valley present in the effective potential depicted in Fig. \ref{fig:potential} created by the reflective barrier at the center. They may yield non-negligible contributions to the image of the object, thus allowing to probe deeper the critical curve while also boosting the contributions of the lower-order rings to the image. These modes are much more sensitive to the features of the background geometry than the other emissions.
\end{itemize}


\begin{figure}[t!]
\includegraphics[width=8.6cm,height=8cm]{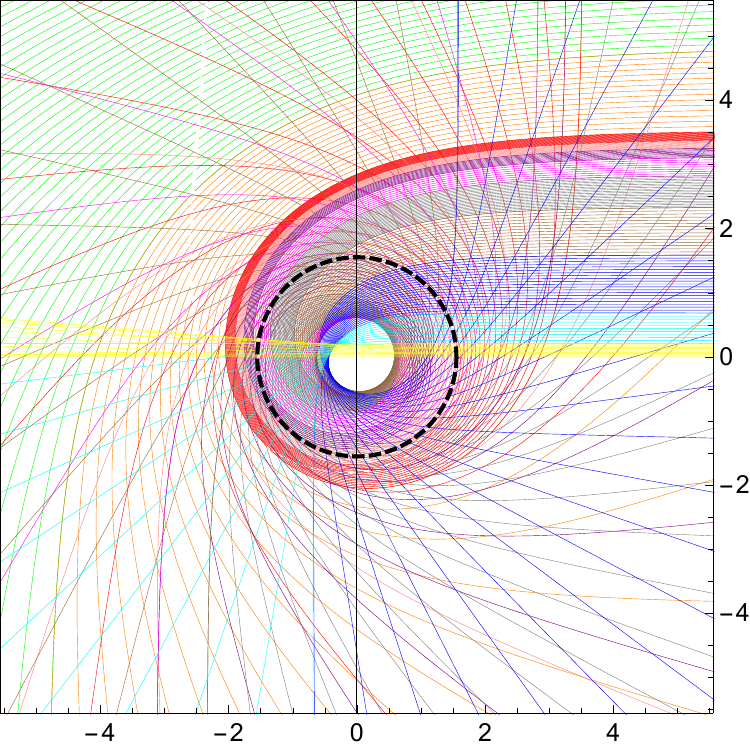}
\caption{Ray-tracing of the modes associated to the direct emission $m=1$ (green, cyan and yellow), the lower-order modes $m=2$ (orange and blue) and $m=3$ (red and brown), and the higher-order modes $m=4$ (gray), $m=5$ (magenta), $m=6$ (purple), $m\geq 7$ (pink). The black dashed circumference corresponds to the critical curve, which is the major force driving these trajectories. Beyond the direct emission ($m=1$), the modes with $m=2,3$ are relevant for the total luminosity both above and below the critical curve, while those with $m=4,5,6$ are non-negligible only below of it. Modes with $m \geq 7$ are too demagnified and can be dismissed from our subsequent analysis.}
\label{fig:raytrac}
\end{figure}

We next implement a 
numerical ray-tracing code to classify the space of parameters corresponding to the direct emission as well as those corresponding to the lower-order and higher-order modes, cutting our integrations at the $m=7$ mode. The corresponding results are depicted in Fig. \ref{fig:raytrac}, where bunches of curves with the same color belong to the same mode. We first notice that approaching to the critical curve (depicted as a black dashed circumference in this plot) by above $b>b_c$ and by below $b<b_c$, has large differences in the contribution of the corresponding modes to the luminosity of the object.
Indeed, for $b>b_c$ we neatly identify the direct ($m=1$) emission in green, followed by the lower-order trajectories $m=2$ (orange) and $m=3$ (red), while contributions to the total luminosity from higher-order modes with $m>3$ will be so demagnified that they can be removed from our analysis. As for $b<b_c$ we identify the relevant contributions by $m=1$ (yellow, cyan), $m=2$ (blue), $m=3$ (brown), $m=4$ (gray), $m=5$ (magenta), $m=6$ (purple) and $m=7$ (purple)

\begin{figure}[t!]
\includegraphics[width=8.0cm,height=5.5cm]{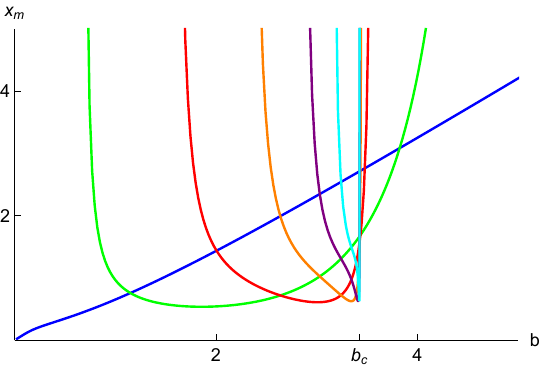}
\caption{The transfer function $r_m$ as a function of $b$ for the trajectories appearing in the ray-tracing plot (\ref{fig:raytrac}): $m=1$ (direct emission, blue), $m=2$ (green), $m=3$ (red), $m=4$ (orange), $m=5$ (purple), $m=6$ (cyan), and $m=7$ (pink). In this figure, $b_c \approx 3.4263$ denotes the location of the critical curve, which is much more reduced than in its Schwarzschild counterpart, $b_c=3\sqrt{3} \approx 5.1962$.}
\label{fig:transferfunction}
\end{figure}

From our analysis below on the total luminosity of the image, we find that only the higher-order modes $m=4,5,6$ will contribute non-negligibly to it, while those modes with $m \geq 7$ can be safely neglected. Intuitively, this enhance on the contribution of the higher-order modes is a direct consequence of the presence of the potential well inside the critical curve, and which can be further supported on the  grounds of the behavior of the so-called ``transfer function", $r_m(b)$. The latter accounts for the location of the point of the equatorial plane (i.e., of the disk) a given light ray with impact factor $b$ will touch on its winding around the SV solution, and whose slope encodes the degree of demagnification of the image associated to a given mode $m$. This transfer function is depicted in Fig. \ref{fig:transferfunction} up to the mode $m=7$ for graphical comparison. Clearly, the direct emission, with its almost constant (unit) slope, dominates the image at every impact parameter value, being insensitive to the existence of the critical curve, which is in agreement with the fact that it reflects source-dependent features rather than probing the background geometry.

The sequence of lower-order and higher-order modes contributes non-negligibly to the luminosity within a (decreasing with higher $m$) range of impact parameter values whose outer edge is roughly located near the critical curve $b=b_c$. As compared to the usual structure of the Schwarzschild solution, the domain of existence of the transfer function for both kinds of modes is greatly enhanced in the inner region of the critical curve due to the presence of the effective potential barrier discussed above, giving comparatively more weight to the lower-order modes and a non-negligible role to the higher-order ones. Indeed, in the Schwarzschild case every transfer function associated to both a lower-order as well as a higher-order mode meets its end shortly after crossing the $b \leq b_c$ region due to the effective potential becoming negative, while in the present case the divergence of the potential at the center allows to extend the bottom end of every mode much deeper inside the critical curve.

The mode $m=1$ corresponding to the inner direct emission (two sets of trajectories depicted in yellow and cyan in Fig. \ref{fig:raytrac}, for reasons that shall be clear below) is deemed of further comment. In a typical black hole space-time, those trajectories with small enough $b<b_c$ do not intersect the equatorial plane even once, defining what is called {\it the inner shadow} \cite{Chael:2021rjo}, corresponding to $n<1/2$. This is so because such light rays intersect the event horizon of the black hole without finding the accretion disk on their path, and therefore the inner shadow, $b=b_{is}$, defines the brightness depression of a black hole no matter the emission properties of the geometrically thin accretion disk. However, such an inner shadow may be missing for a horizonless compact object. Examples of this are traversable wormholes  (i.e. horizonless), in which the inner shadow limit may be defined by the size of its throat (though for $b<b_{is}$ light rays  traveling from the other side of the throat could still reach the observer on our side \cite{Olmo:2021piq}), or by objects without a critical curve such as gravastar/boson star/Proca stars \cite{Herdeiro:2021lwl,Rosa:2022tfv}, where no multi-ring structure contributing to the direct image is present. For the SV objects considered here, their inner shadow will be mainly determined by the position of the inner edge of the disk, since the emission profiles chosen in this work will be truncated (and take their maximum values) at such a surface, smoothly decreasing onwards. This means that when such an edge is allowed to extend close enough to the center of the object, the latter can become shadowless, as shall be verified later on once we run our luminosity code.

There is yet another interesting feature worth of comment: for small enough impact parameter, $b \lesssim 0.2$ (yellow trajectories in Fig. \ref{fig:raytrac}), a new feature for this SV object emerges, consisting on a slightly divergent lens effect (rather than convergent) on the corresponding light rays. The closest to the center, the weaker the deviation is, whereas the maximum divergent lens effect occurs around $b \approx 0.2$, and is immediately followed by the standard convergent behaviour for larger impact parameters. This effect can be traced back to the effective energy sources that - within the context of GR - generate this geometry. Indeed, one can verify that the stress energy tensor of this line element can be mimicked by a nonlinear theory of electrodynamics whose (positive) energy density is peaked at $r_{max}=l/4=0.1875$, and has negative pressures with minima at $r_{rad}=l/4=0.1875$ (radial component) and at $r_{ang}=0.13485$ (angular components). Typically, in wormhole \cite{VisserBook} and black bounce space-times \cite{Bronnikov:2021uta} negative pressures are required in order to prevent the focusing of geodesics via a local repulsive gravity effect it produces, and we similarly interpret this shell of negative pressure as responsible for the divergent lens effect. Since the center has vanishing energy and pressures, the lowest impact factors experience little or no deflection at all.

\subsection{Emission from the thin accretion disk}

In order to implement our numerical simulation of the optical appearances of the SV solution above, we first summarize the assumptions and simplifications upon which our accretion disk model is built. First of all, we recall our assumption of an optically thin disk, which means that a photon emitted by the disk and crossing again the disk after winding one or more times around the object will not be reabsorbed. This allows light rays associated to the lower-order and higher-order modes to arrive to the observer's screen and to add their luminosity to that of the direct emission (which does not cross the disk again). Secondly, and  in order to actually be able to see this multi-ring structure, we assume the disk to be infinitesimally thin, and furthermore conveniently located on the equatorial plane used in the ray-tracing. Finally, we assume a  monochromatic emission with specific intensity $I_{\nu}^{em}=I(r)$ and zero absorptivity. Under these simplifications, the Boltzmann equation of radiative transfer \cite{Gold:2020iql} tells us the conservation of the quantity $I_{\nu}^{em}/\nu^3$ along a given photon's trajectory, which in the observer's frame with an associated frequency $\nu'$, and according to the line element (\ref{eq:motion}), it is translated into $I_{\nu'}^{ob}=A^{3/2}(r)I(r)$. Then, by integrating over the full spectrum of observed frequencies $I^{ob}=\int d\nu'I_{\nu'}$, the total observed intensity is obtained $I^{ob}=A^2(r)I(r)$. However, we need to incorporate into this formula the actual values of the disk radii at which light rays with given impact parameter hit, an info nicely encoded in the transfer function already depicted in Fig. \ref{fig:transferfunction}. Therefore, adding all the possible intersections with the disk for every light trajectory and cutting the sequence at the $m=6$ mode in order to neglect additional higher-order contributions\footnote{Recall that such higher orders have no visible effects on the optical appearance images, as can be easily inferred from Fig.~\ref{fig:transferfunction}. Neglecting them also shortens computation times, which is a heavy practical limitation.}, the formula that captures all these elements is finally given by:
\begin{equation} \label{eq:Iob}
I^{ob}=\sum_{m=1}^{6} A^2(r)I(r)\Big\vert_{r=r_m(b)} \ .
\end{equation}

The next step in our analysis is to set an emission profile for $I(r)$ in the effective region of the (infinitely thin) disk. We are modelling such a profile by truncating it at the inner edge of the disk, $r_{ie}$, where it actually takes its maximum value, and smoothly decreasing outwards until asymptotic infinity (so that the outer edge of the disk is assumed to be infinitely far away) with a given radial decay. Obviously, both aspects are subject to heavy astrophysical modelling and uncertainties, which we will not discuss here (see instead the detailed analysis of  \cite{Gold:2020iql} for the inferred disk's properties according to magnetohydrodynamic simulations). To simplify the analysis of this aspect, typically in the literature different decay profiles for the emission are taken \textit{ad hoc} depending on how close to the innermost region of the geometry the inner edge of the disk is, and we ourselves have used this strategy in some previous works \cite{Guerrero:2021ues,Olmo:2021piq}. Here, we somewhat reverse such an approach since we are interested in analyzing how the lower-order and higher-order ring structures get modified as we move the inner edge of the disk closer to the inner regions of the geometry while keeping the decay of the emission profile fixed. To this end, we shall consider three different decay laws and move in discrete steps the inner edge of the disk for each of them, starting from the innermost stable circular orbit for time-like observers, downward until getting to the very center of the object, $r=0$, since the absence of an event horizon does not prevent the accreting material to keep falling all the way down (though the infinite slope of the potential will do it so). The goal of such an approach is to compare the evolution in  size and luminosity of the lower and higher-order photon rings (created by its respective modes), for a fixed emission model while also allowing us to compare the optical appearance of the object at fixed inner edge but different emission profiles.

\begin{figure}[t!]
\includegraphics[width=8.0cm,height=5.5cm]{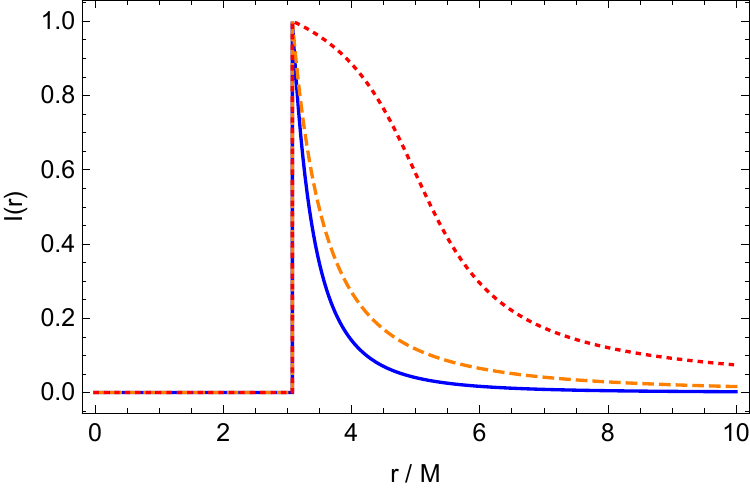}
\caption{The three (normalized) emission profiles (\ref{eq:I1}), (\ref{eq:I2}), and (\ref{eq:I3}) [solid blue, dashed orange, and dotted red, respectively] when $r_{ie}=r_{ISCO} \approx 3.0878$, where we appreciate their different decay features with radial distance. }
\label{fig:emission}
\end{figure}

\begin{figure}[t!]
\includegraphics[width=8.0cm,height=5.3cm]{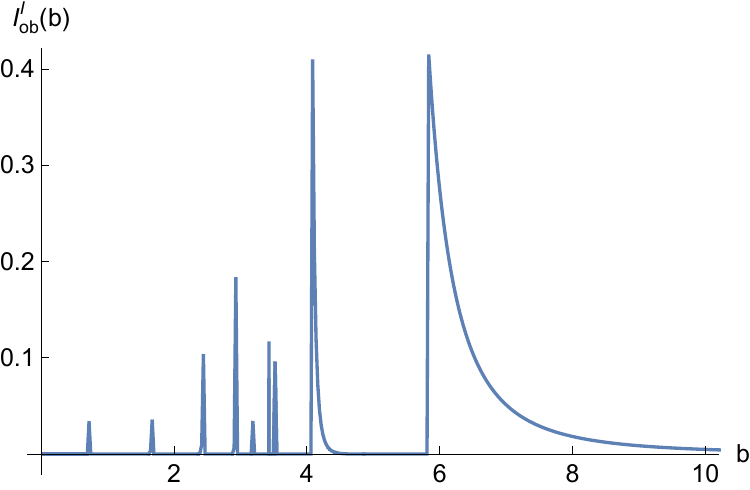}
\caption{The observed intensity  (\ref{eq:Iob})  as function of the impact parameter $b$, where we have taken for convenience the emission model (\ref{eq:I2}) with $r_{ie}=5$ in order to clearly see the  peaks associated to the (eight) narrow rings, and where the outermost curve corresponds to the direct emission, which is much more spread than those of the multi-rings.  }
\label{fig:Iob}
\end{figure}

To be more specific, for the sake of our simulations we shall consider the following three emission models
\begin{eqnarray}
I_{1}(r)&=&\left\lbrace\begin{array}{l}\frac{1}{(r-(r_{ie}-1))^3} \hspace{0.2cm} \text{if} \hspace{0.1cm} r \geq r_{ie} \\ 0 \hspace{1.8cm} \text{if} \hspace{0.1cm} r < r_{ie} \end{array}\right. \label{eq:I1} \\
I_{2}(r)&=&\left\lbrace\begin{array}{l}\frac{1}{(r-(r_{ie}-1))^2} \hspace{0.2cm} \text{if} \hspace{0.1cm} r \geq r_{ie} \\ 0 \hspace{1.8cm} \text{if} \hspace{0.1cm} r < r_{ie} \end{array}\right. \label{eq:I2}  \\
I_{3}(r)&=&\left\lbrace\begin{array}{l}\frac{\pi/2-\arctan[r-5]}{\pi/2-\arctan[r_{ie}-5]} \hspace{0.2cm} \text{if} \hspace{0.1cm} r \geq r_{ie} \\ 0 \hspace{2.5cm} \text{if} \hspace{0.1cm} r < r_{ie} \end{array}\right. \label{eq:I3}
\end{eqnarray}
Here the intensity is normalized to one at the beginning of the emission corresponding to the location of the inner edge of the disk, $r=r_{ie}$, where the above profiles are truncated. These profiles are depicted in Fig. \ref{fig:emission} in which we set, for convenience, the location of  $r_{ie}$ at the innermost stable circular orbit (ISCO) for time-like observers, which for  the present geometry corresponds to $r_{ISCO} \approx 3.0878$. This will actually be the first surface considered in our analysis of the images of the background geometry, as explained below. With the aim of our discussion below, we also depict in Fig. \ref{fig:Iob} the observed intensity (\ref{eq:Iob}) for emission model (\ref{eq:I2}) when $r_{ie}=5$, the latter choice set in order to clearly distinguish the sequence of peaks associated to lower and higher-order rings.

\begin{figure}[t!]
\includegraphics[width=8.0cm,height=5.3cm]{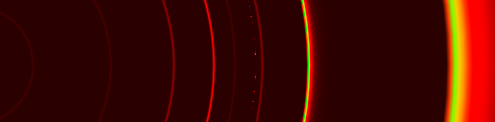}
\caption{The multi-ring structure associated to the observed intensity of Fig. \ref{fig:Iob}. The ring located on the far right corresponds to the direct emission, and inner to it one can see up to eight additional rings.}
\label{fig:cut}
\end{figure}

\begin{figure*}[t!]
\includegraphics[width=5.9cm,height=5.0cm]{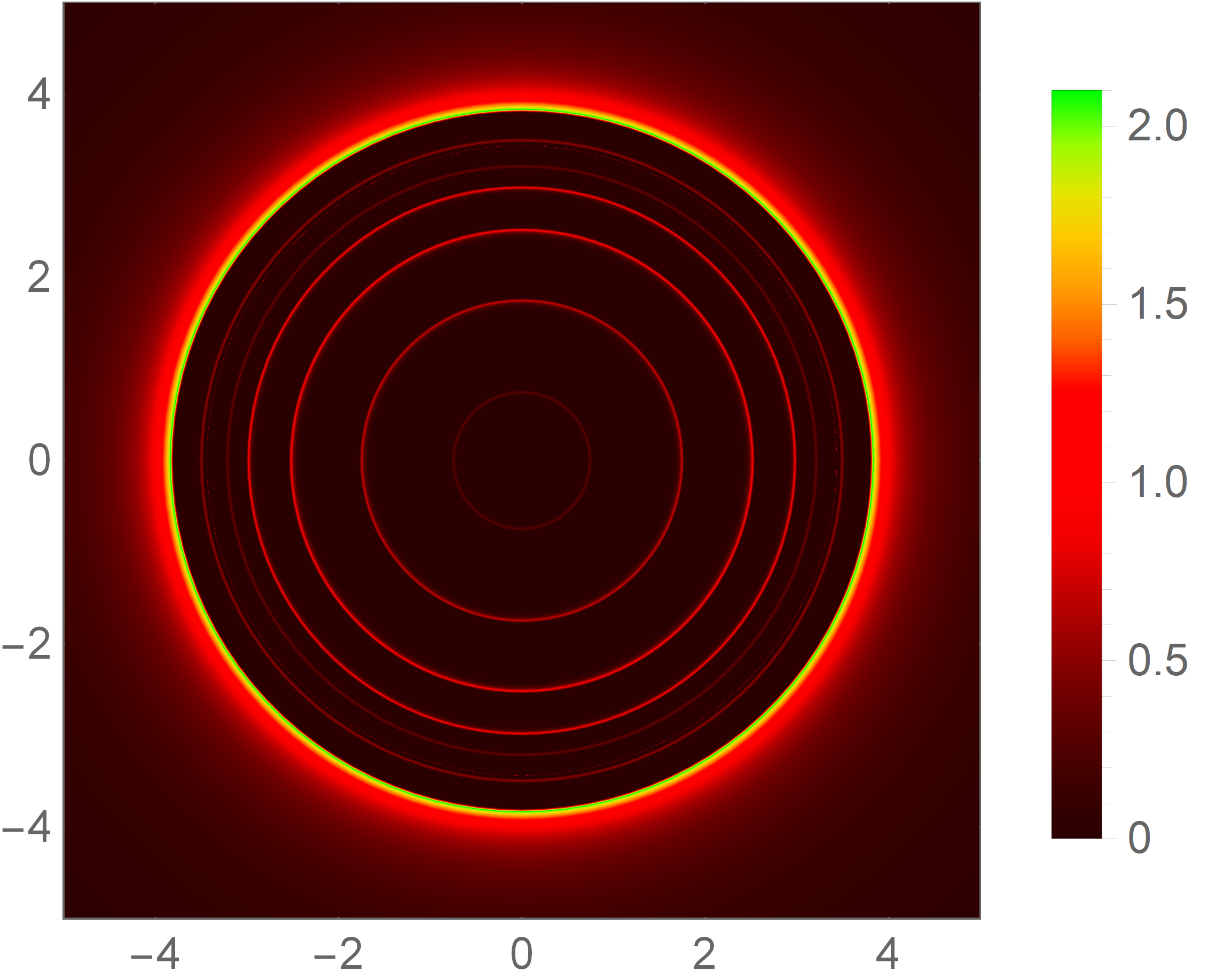}
\includegraphics[width=5.9cm,height=5.0cm]{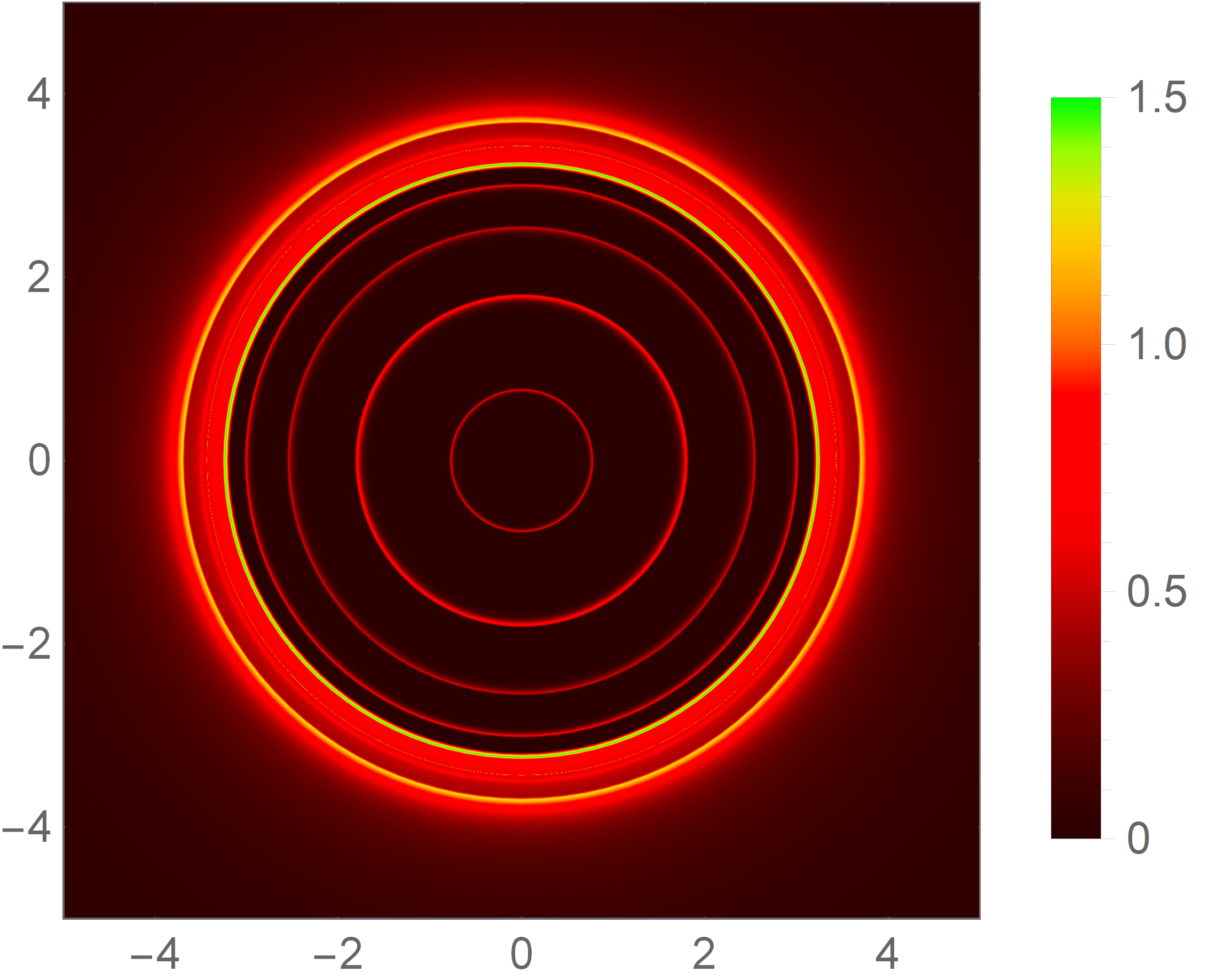}
\includegraphics[width=5.9cm,height=5.0cm]{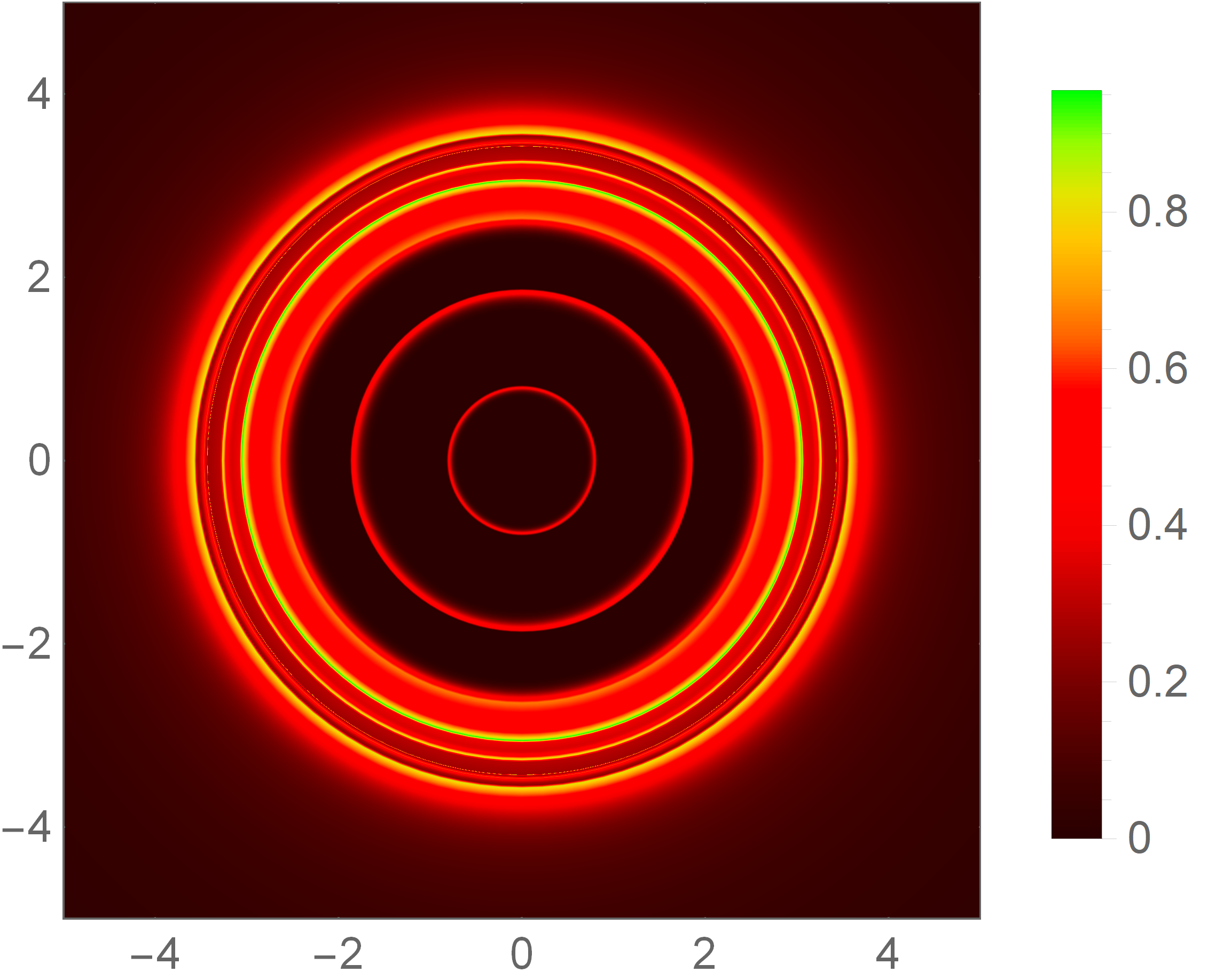}
\includegraphics[width=5.9cm,height=5.0cm]{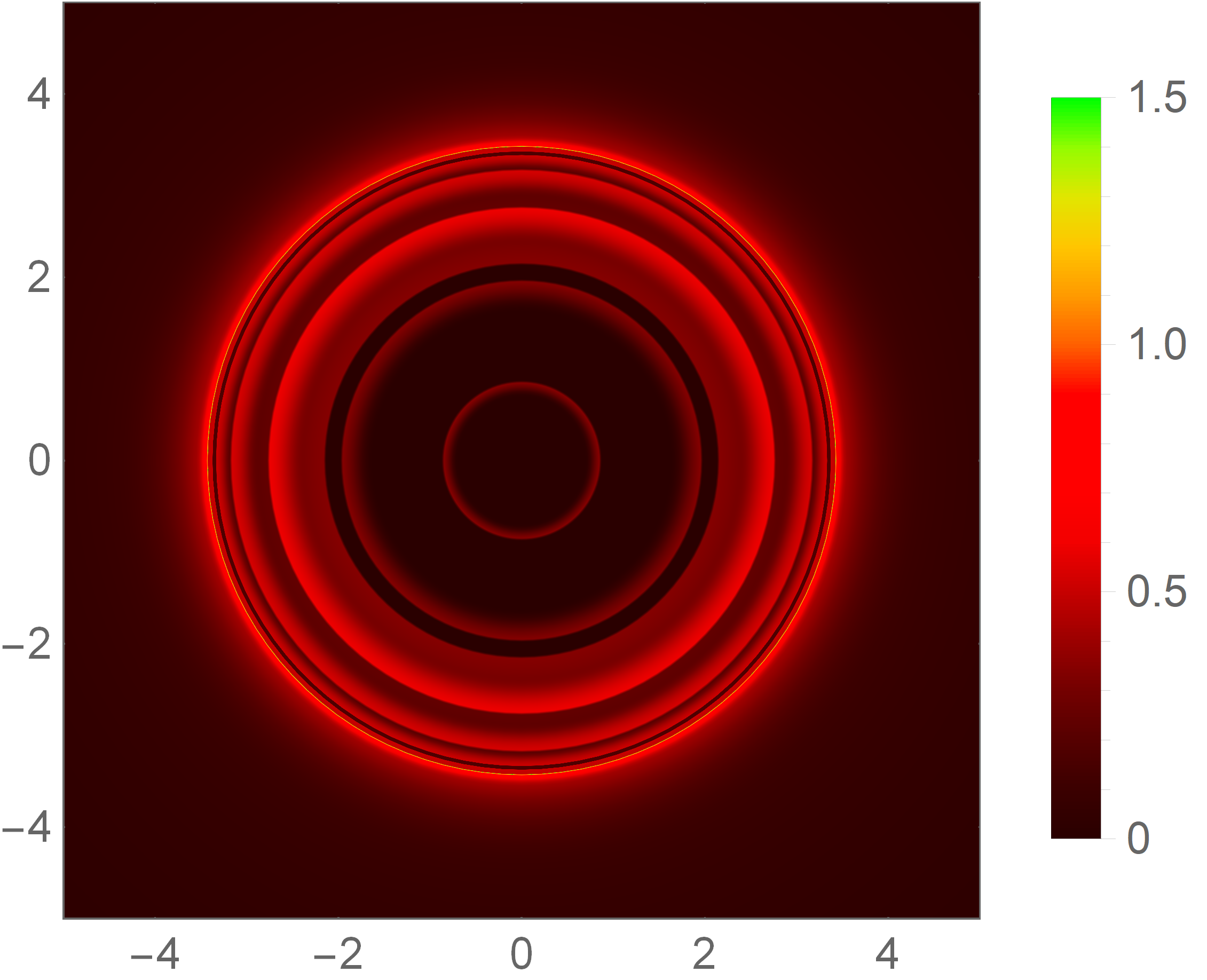}
\includegraphics[width=5.9cm,height=5.0cm]{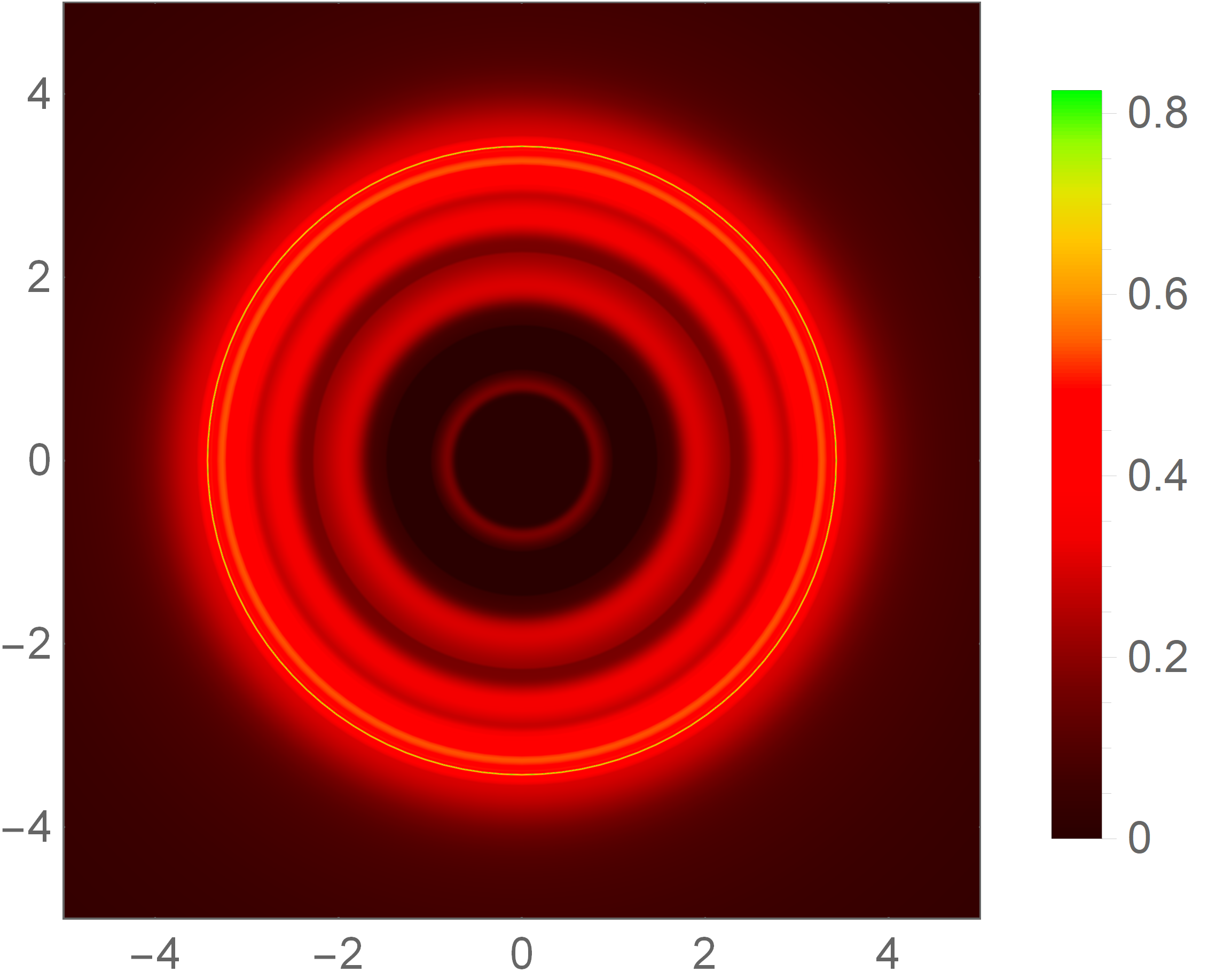}
\includegraphics[width=5.9cm,height=5.0cm]{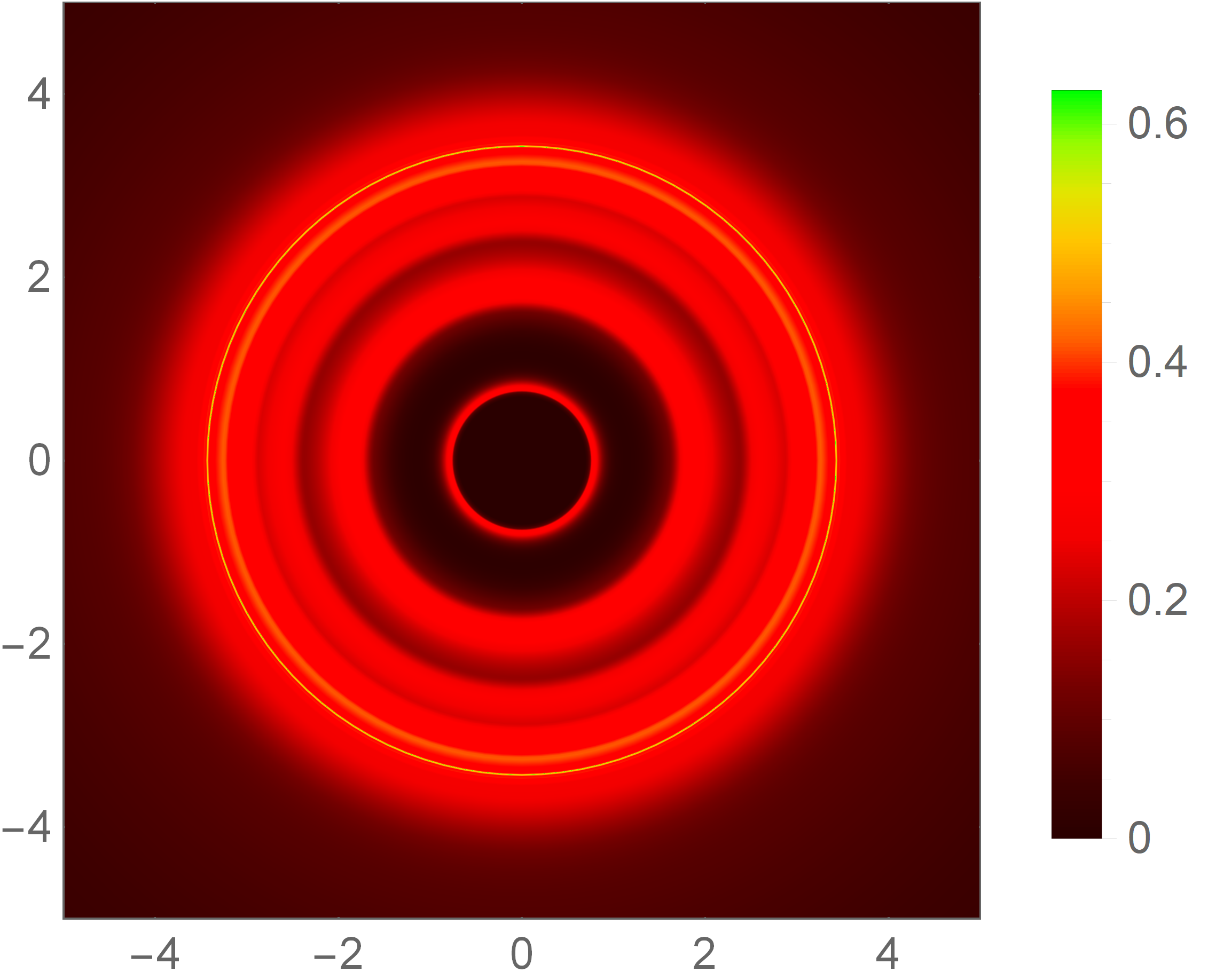}
\includegraphics[width=5.9cm,height=5.0cm]{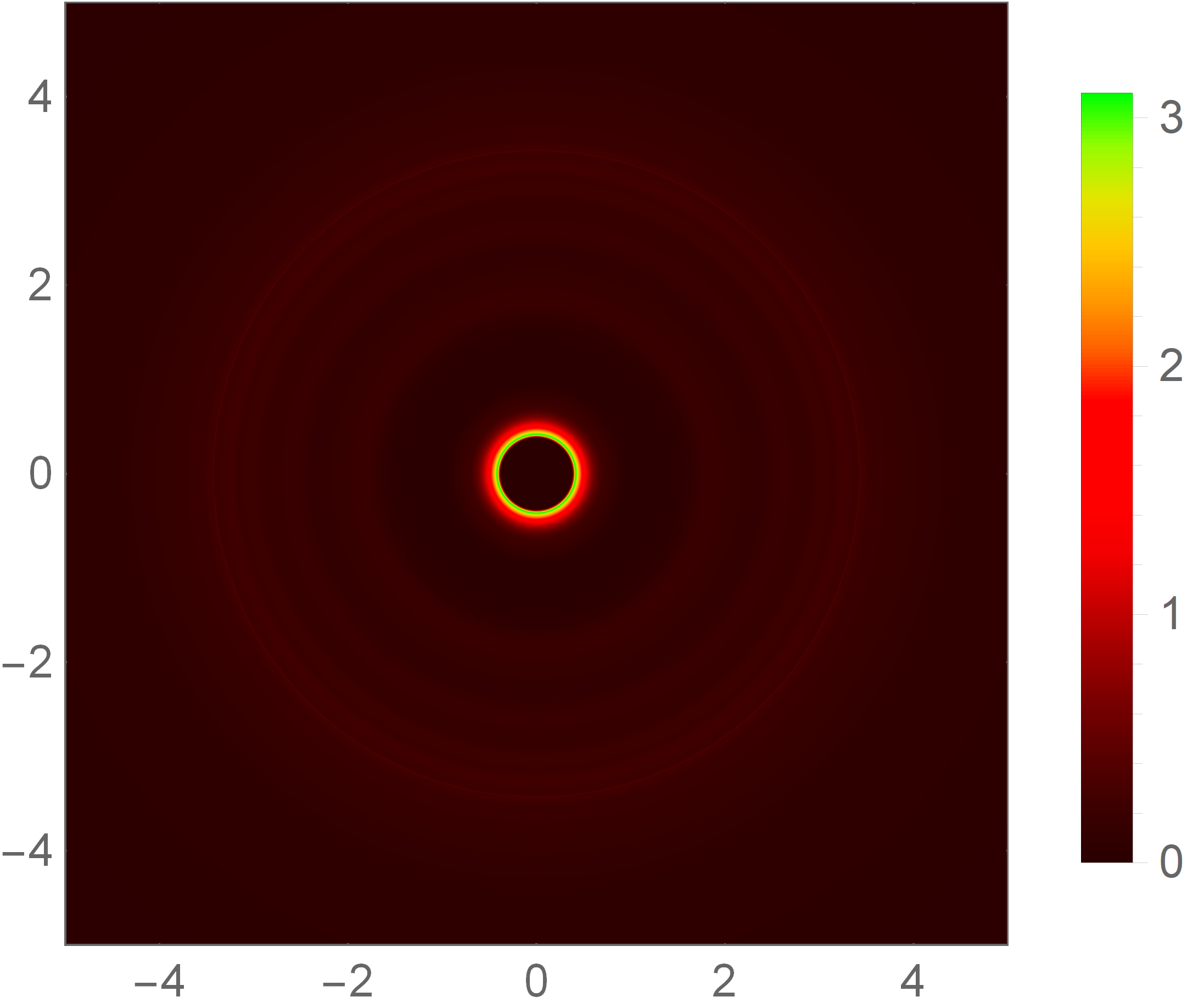}
\includegraphics[width=5.9cm,height=5.0cm]{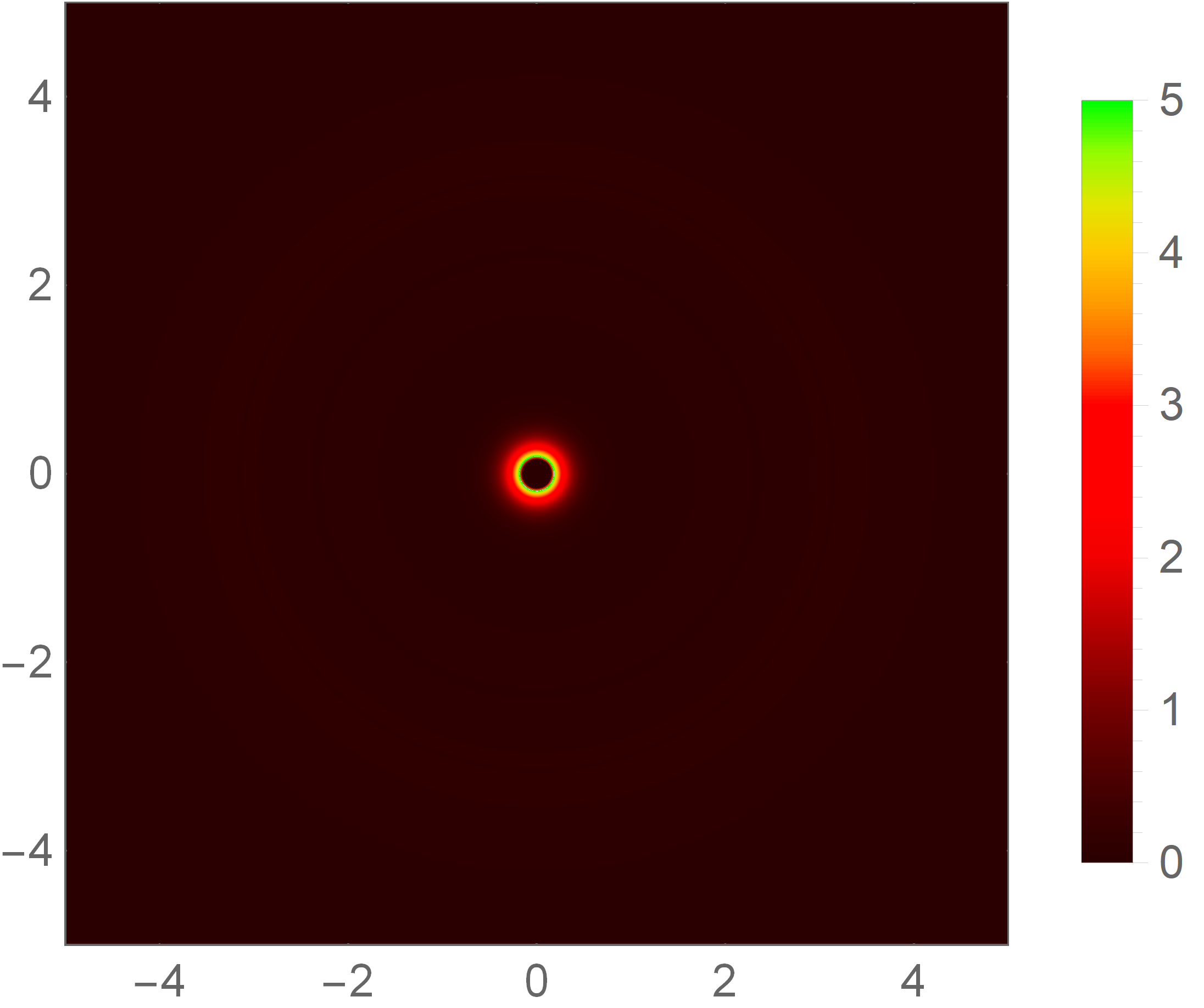}
\includegraphics[width=5.9cm,height=5.0cm]{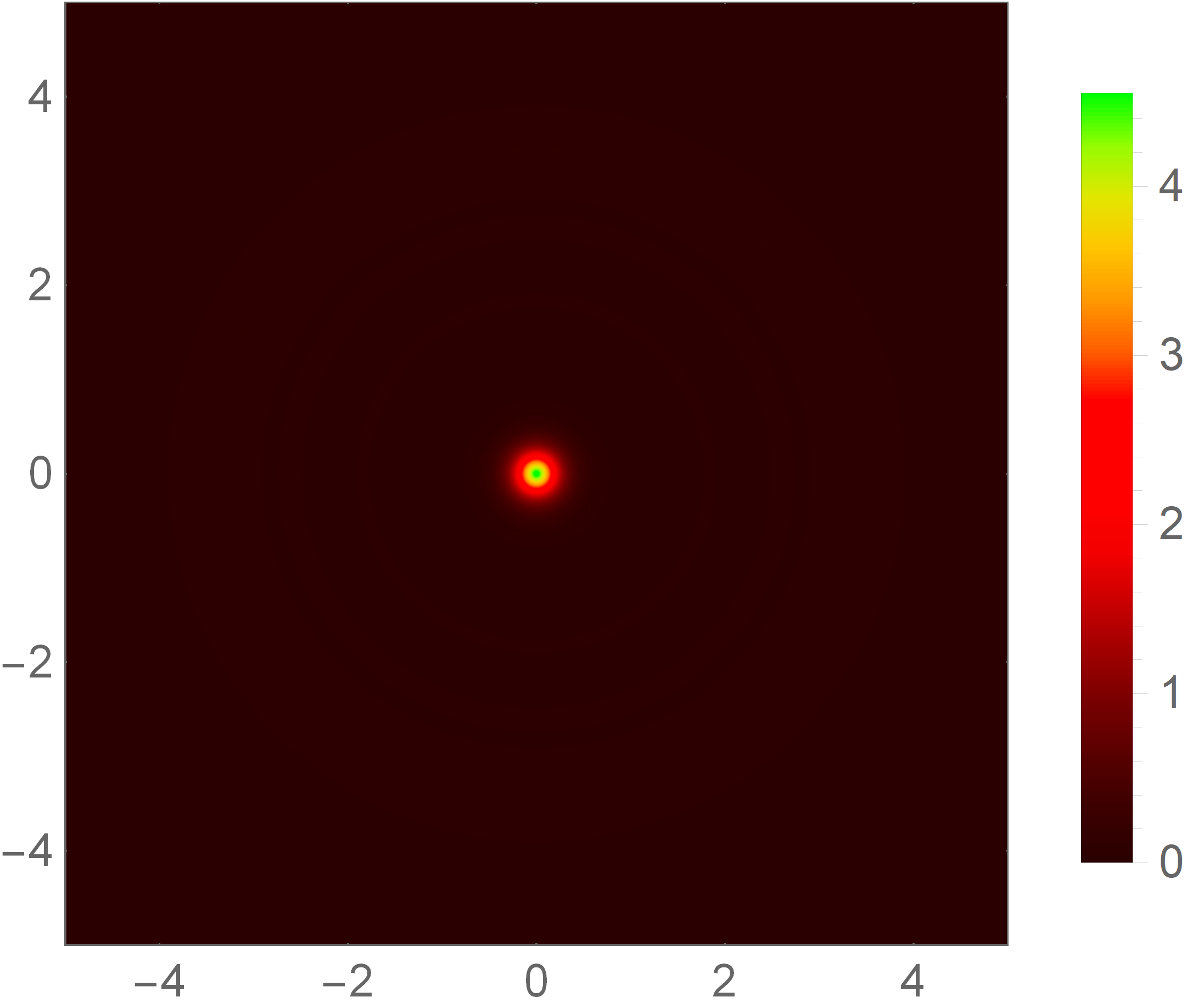}
\caption{The optical appearance of the regular naked SV compact object considered in this work  in the emission model (\ref{eq:I1}) for nine values of the location of its inner edge (from left to right and top to bottom): $r=\{3.0878 (r_{ISCO}),2.5,2.0, 1.5528 (r_{m}), 1.0,0.5,0.30,0.15,0\}$. The structure of lower-order and higher-order photon rings is mostly apparent in the first figure of the sequence, while its superimposition with the direct emission as we move through the sequence produces more complex and hard-to-split contributions of each ring to the image.}
\label{fig:accmod1}
\end{figure*}

\begin{figure*}[t!]
\includegraphics[width=5.9cm,height=5.0cm]{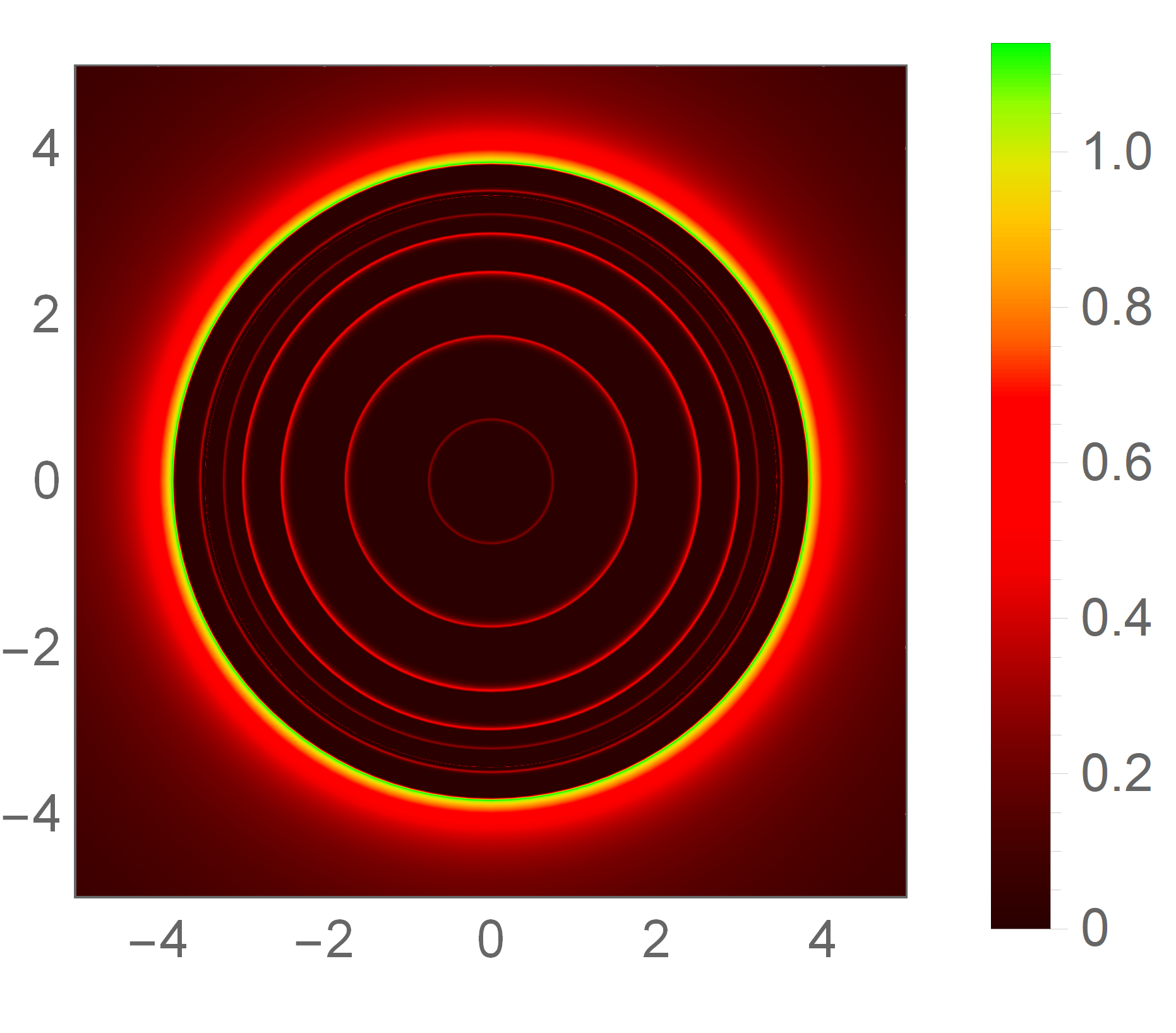}
\includegraphics[width=5.9cm,height=5.0cm]{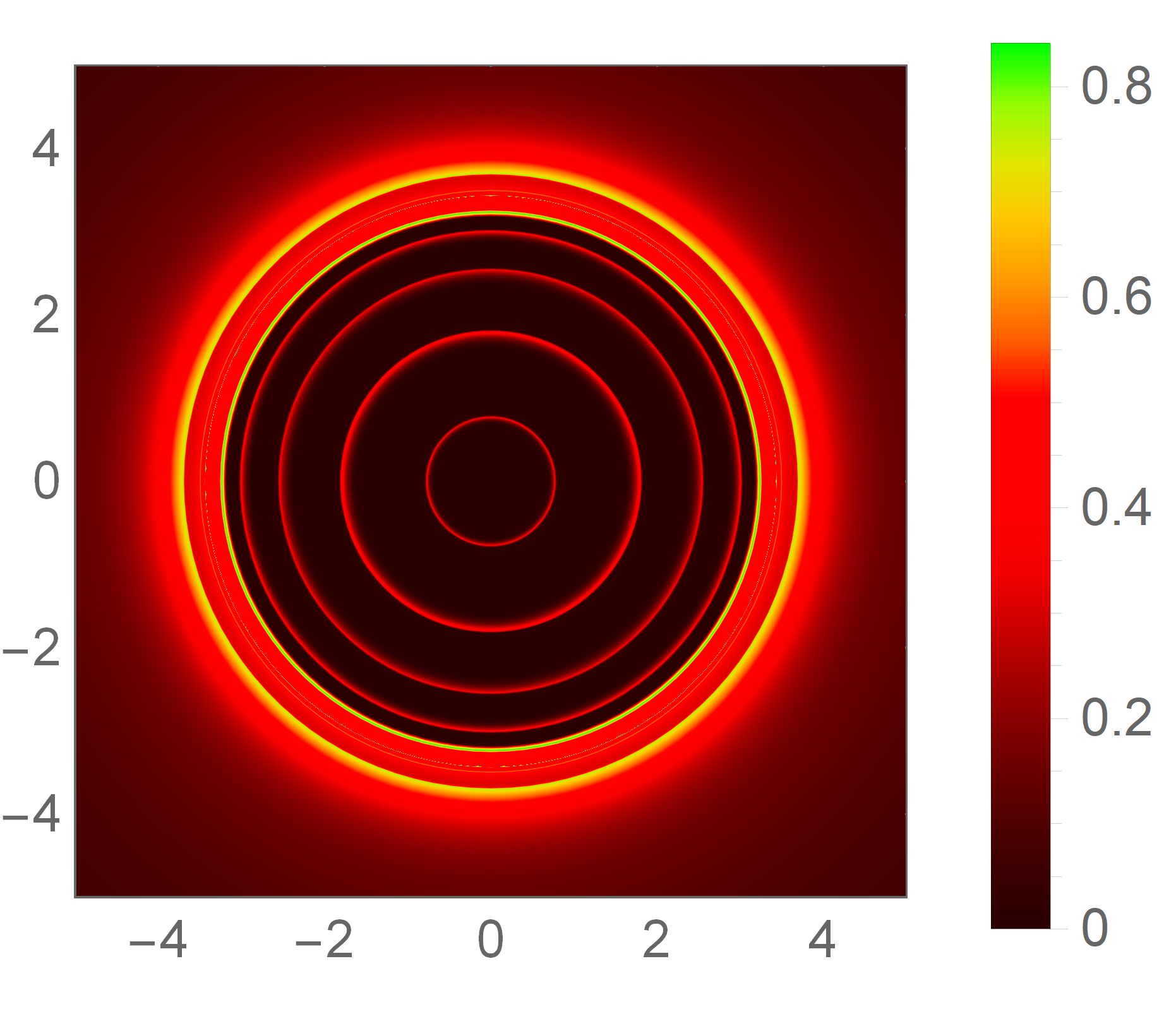}
\includegraphics[width=5.9cm,height=5.0cm]{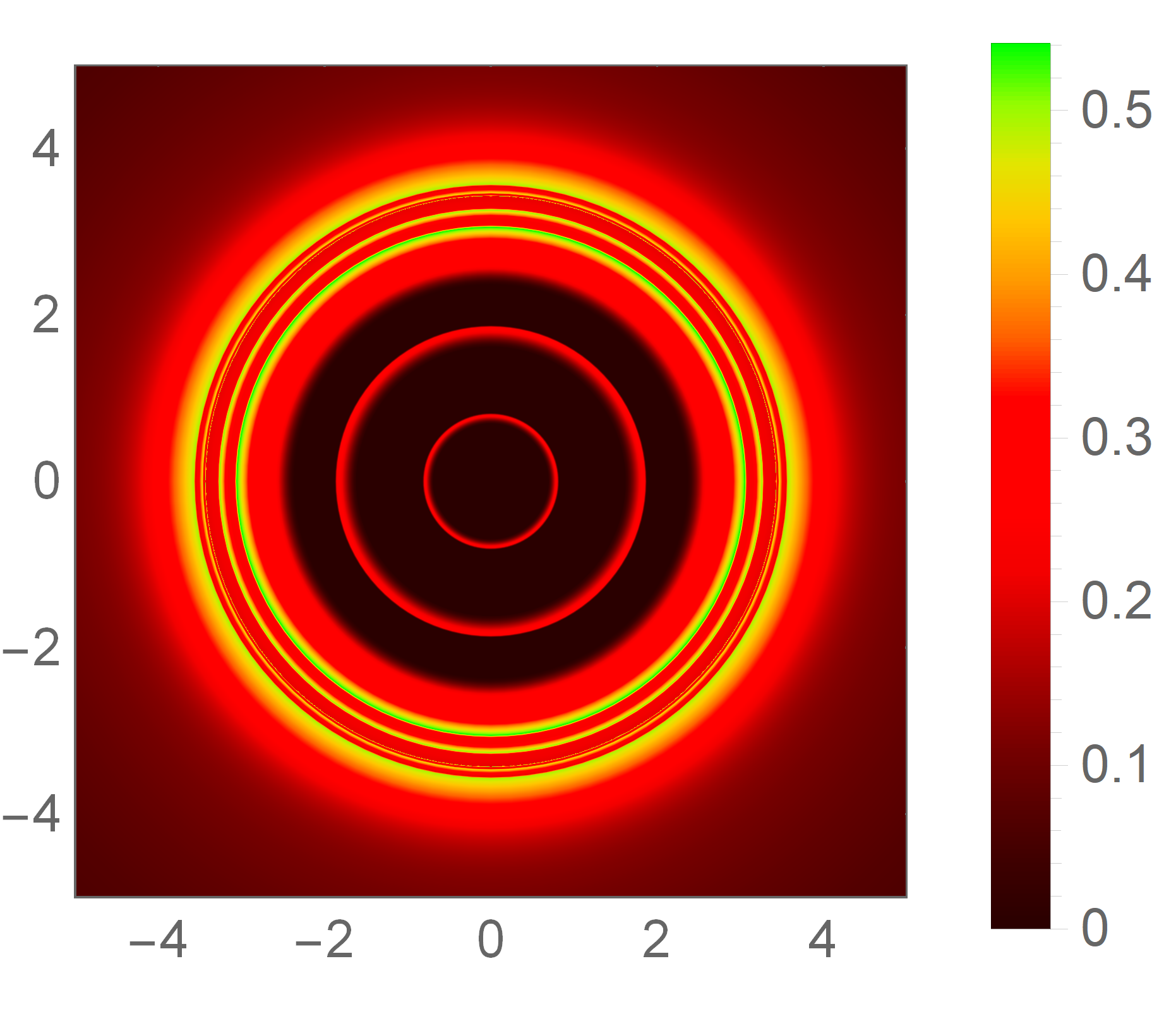}
\includegraphics[width=5.9cm,height=5.0cm]{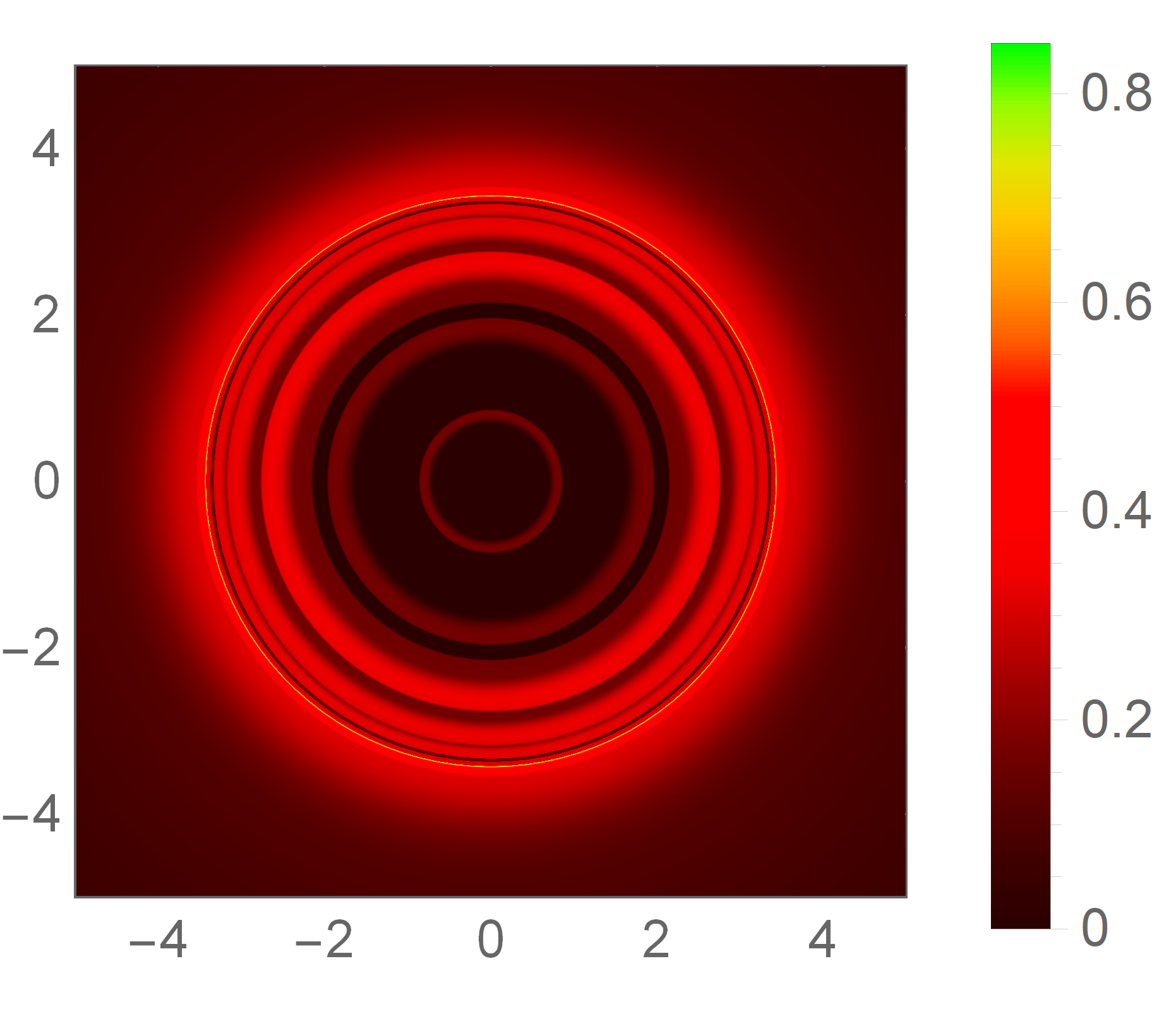}
\includegraphics[width=5.9cm,height=5.0cm]{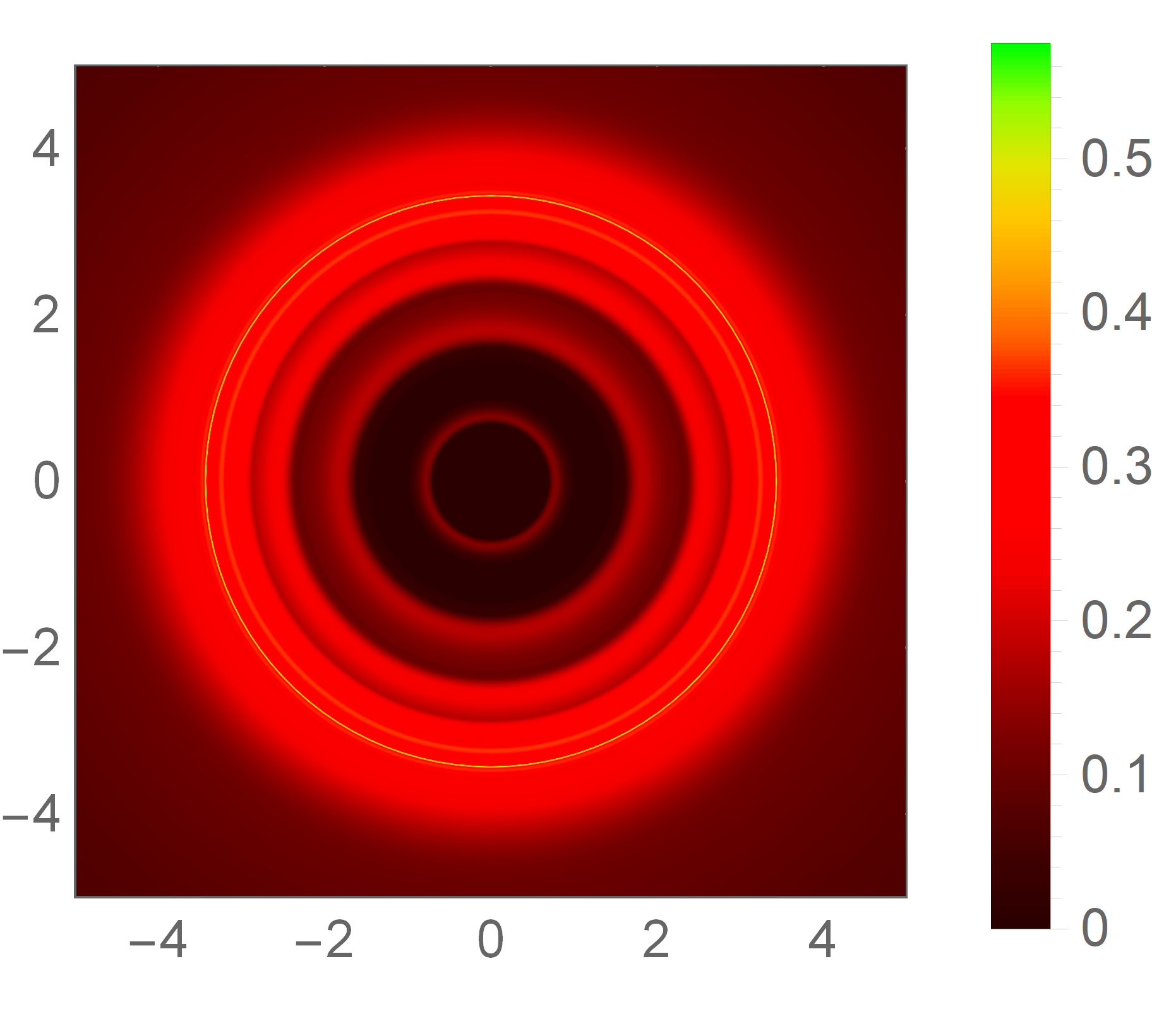}
\includegraphics[width=5.9cm,height=5.0cm]{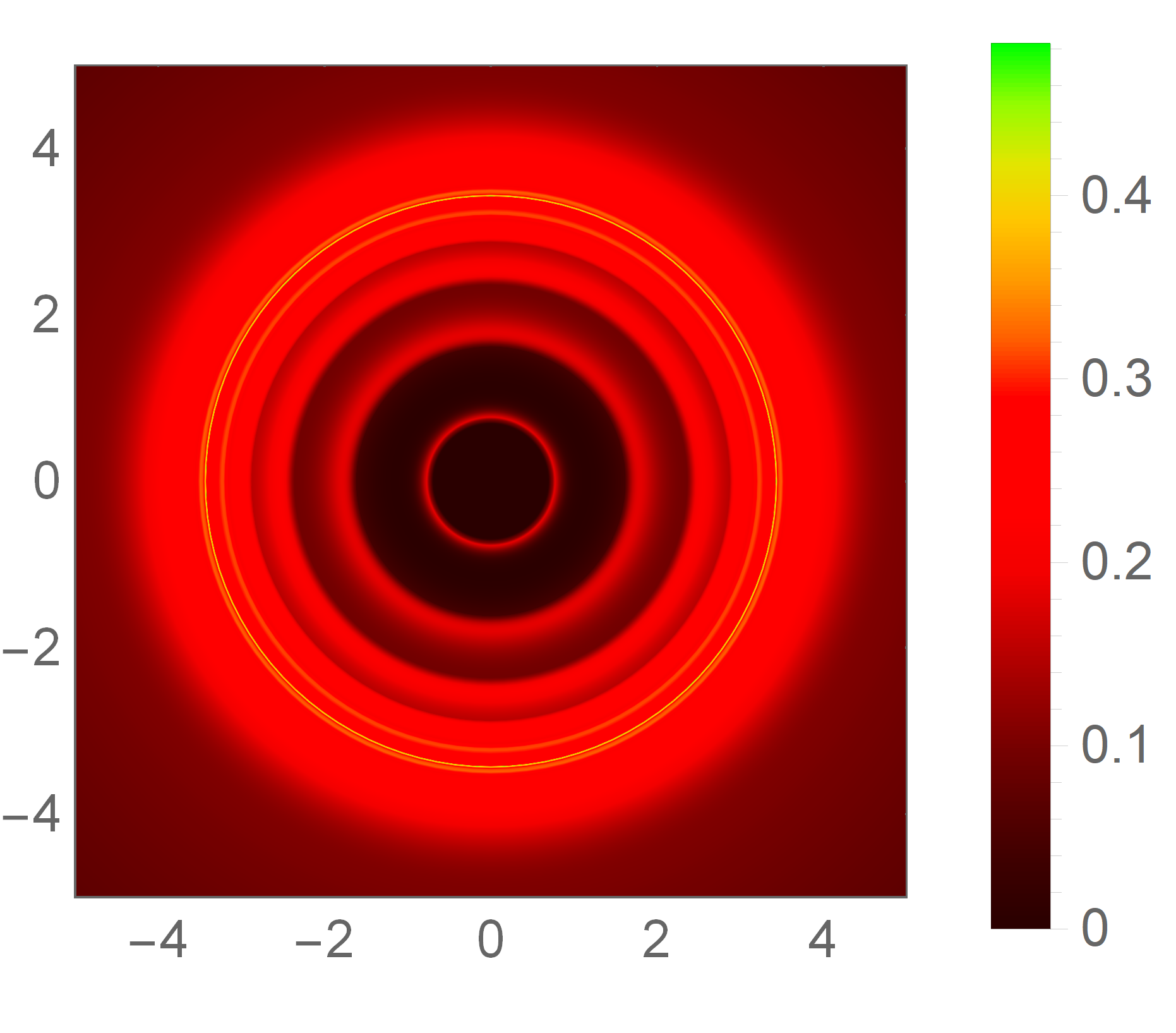}
\includegraphics[width=5.9cm,height=5.0cm]{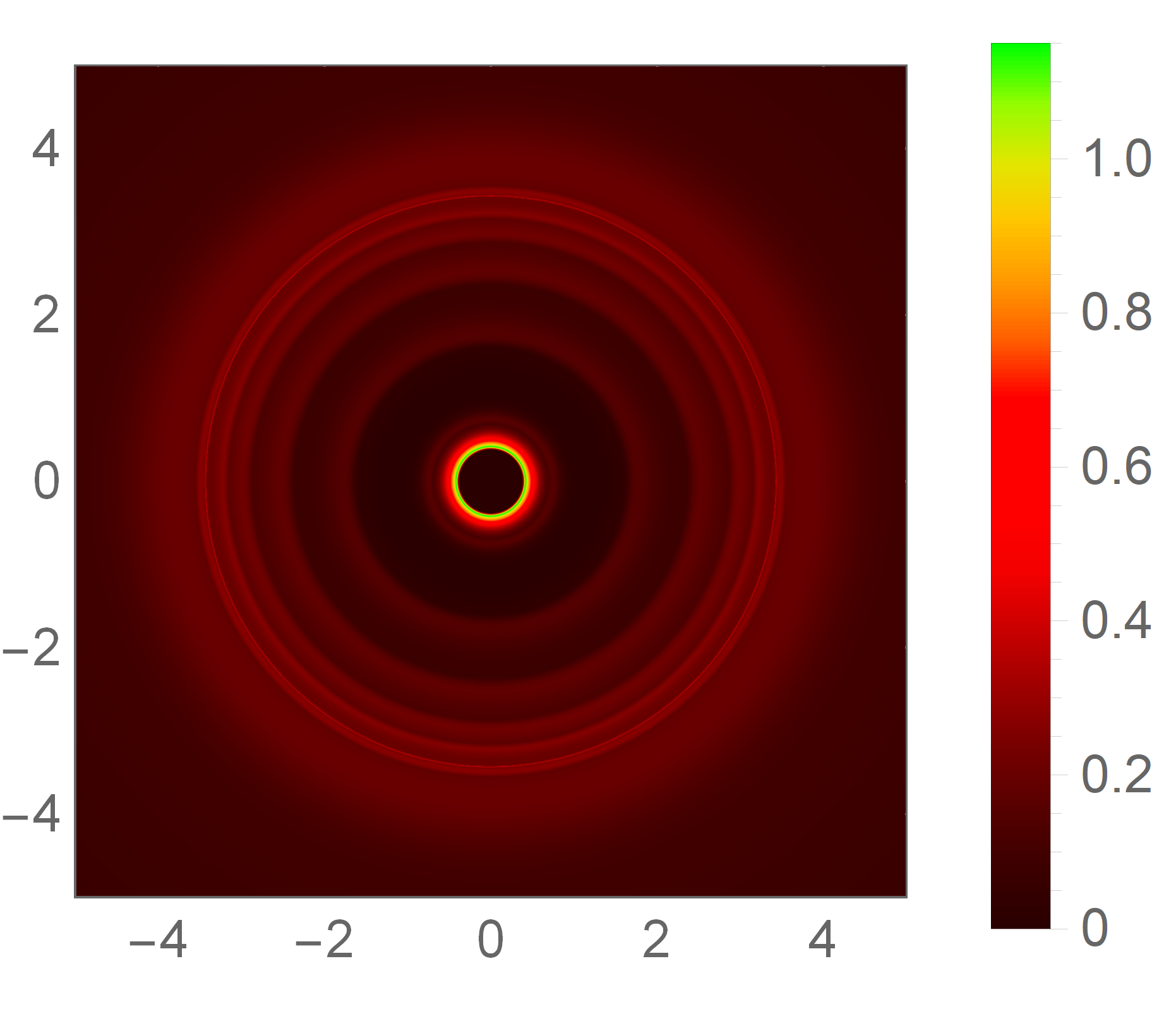}
\includegraphics[width=5.9cm,height=5.0cm]{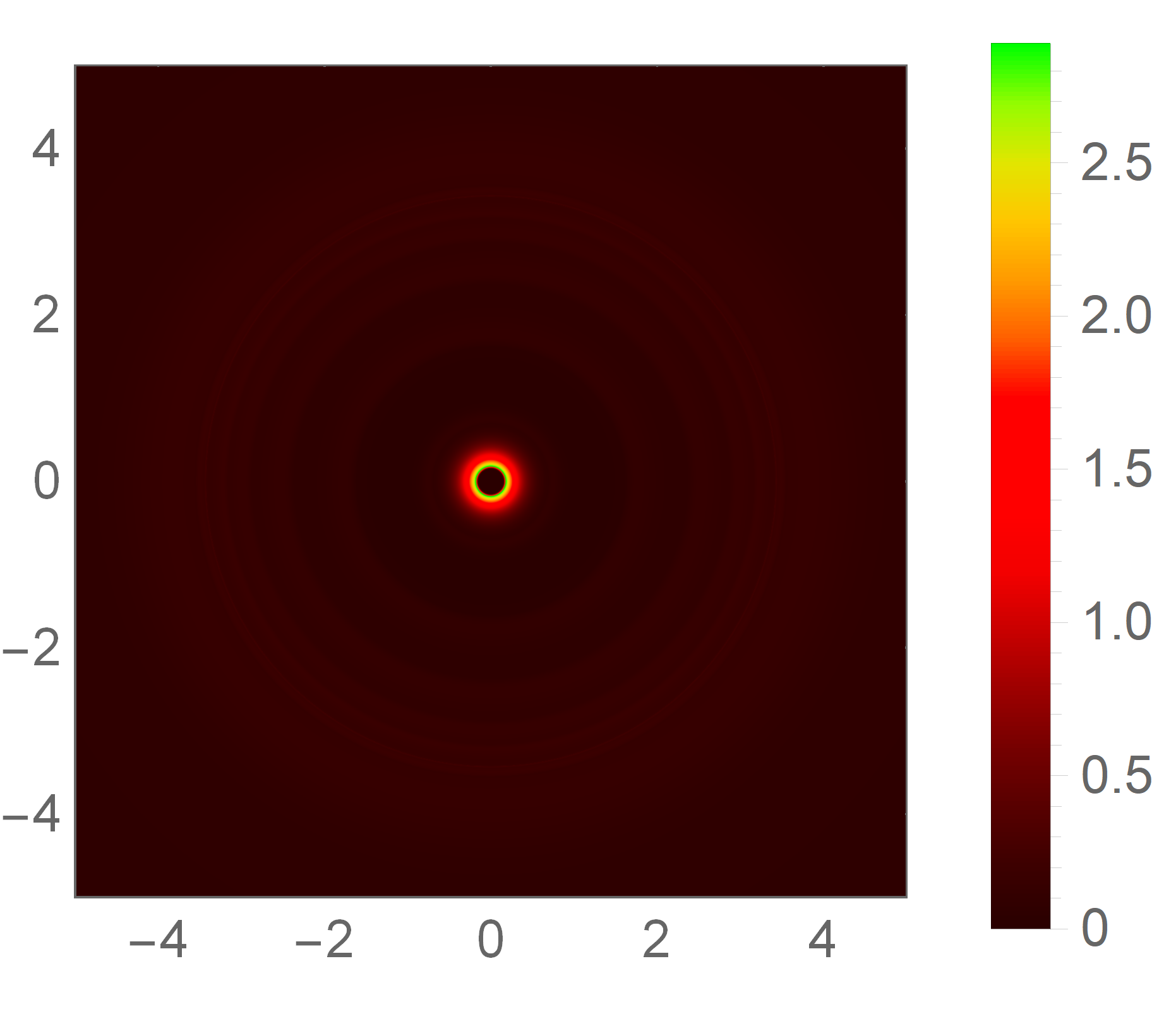}
\includegraphics[width=5.9cm,height=5.0cm]{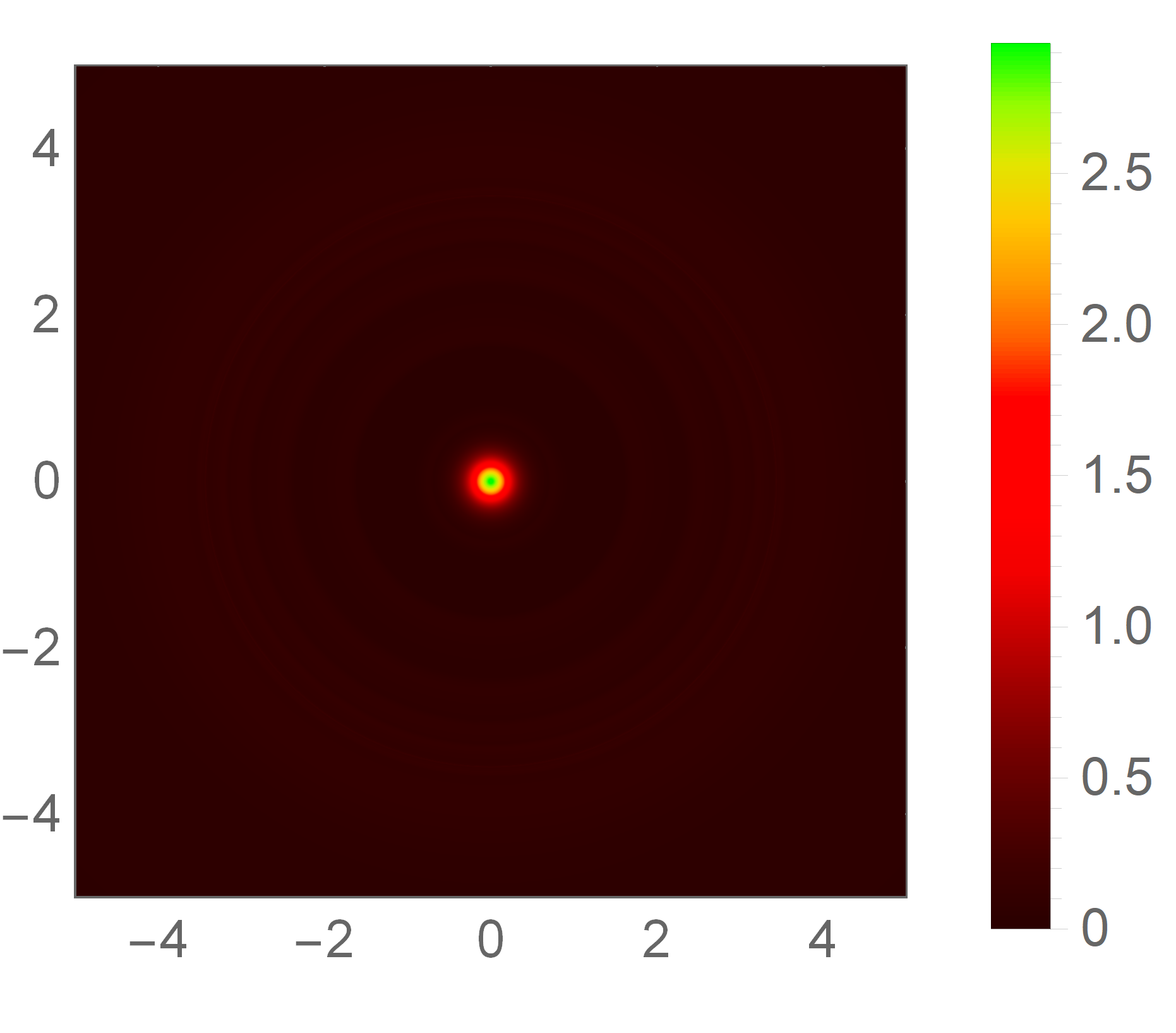}
\caption{The optical appearance of the SV object in the emission model (\ref{eq:I2}). Same notation as in Fig. \ref{fig:accmod1}.}
\label{fig:accmod2}
\end{figure*}

\begin{figure*}[t!]
\includegraphics[width=5.9cm,height=5.0cm]{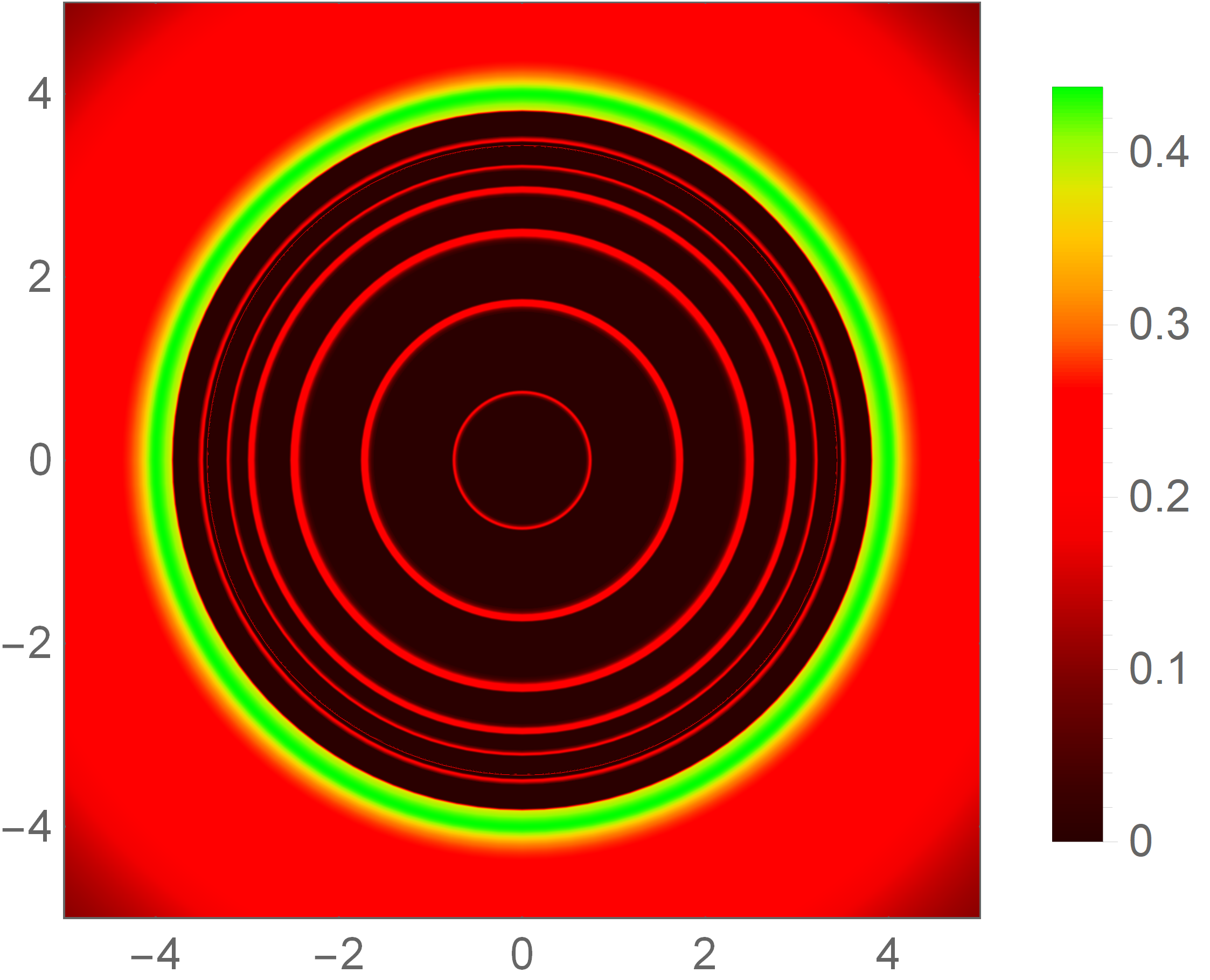}
\includegraphics[width=5.9cm,height=5.0cm]{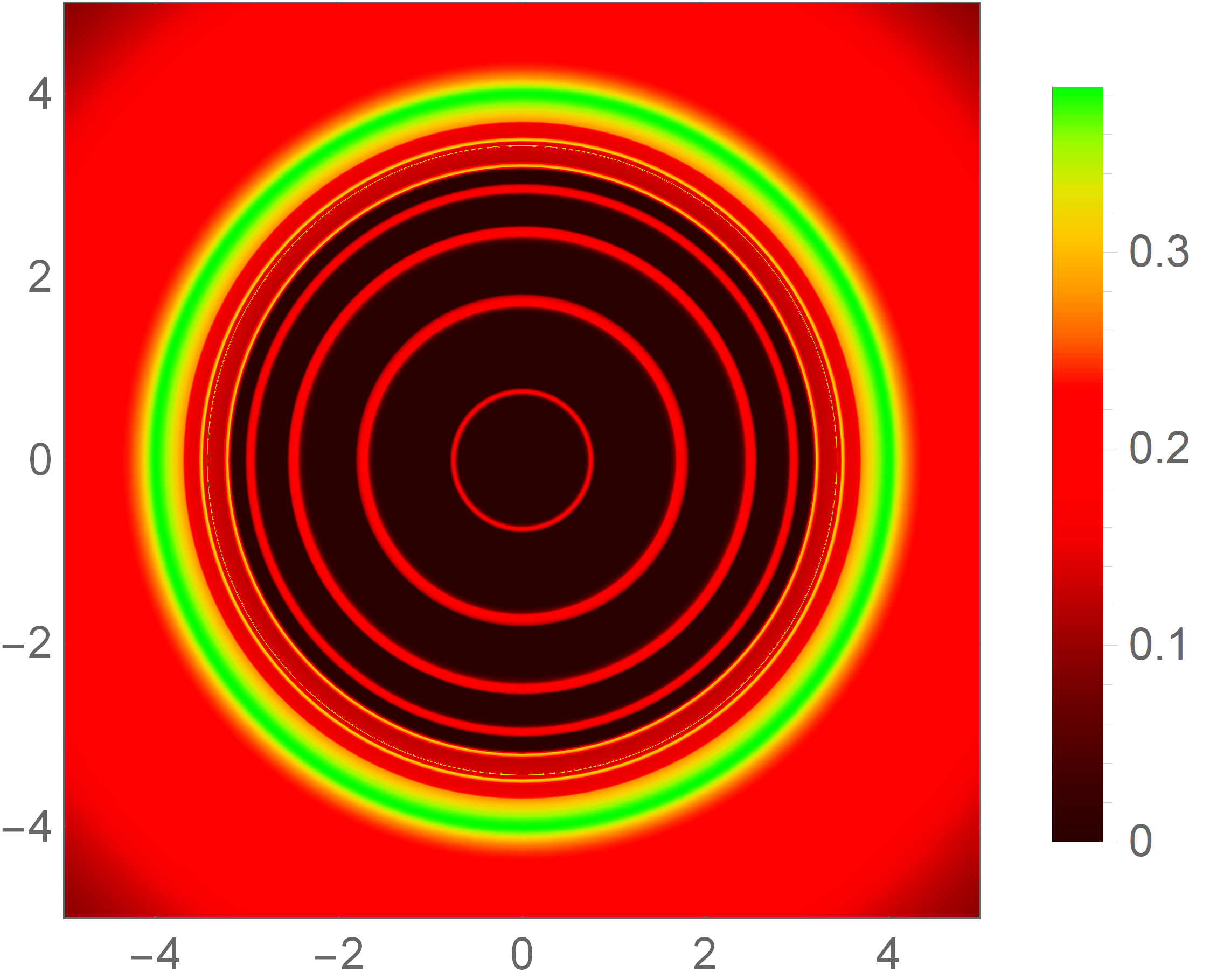}
\includegraphics[width=5.9cm,height=5.0cm]{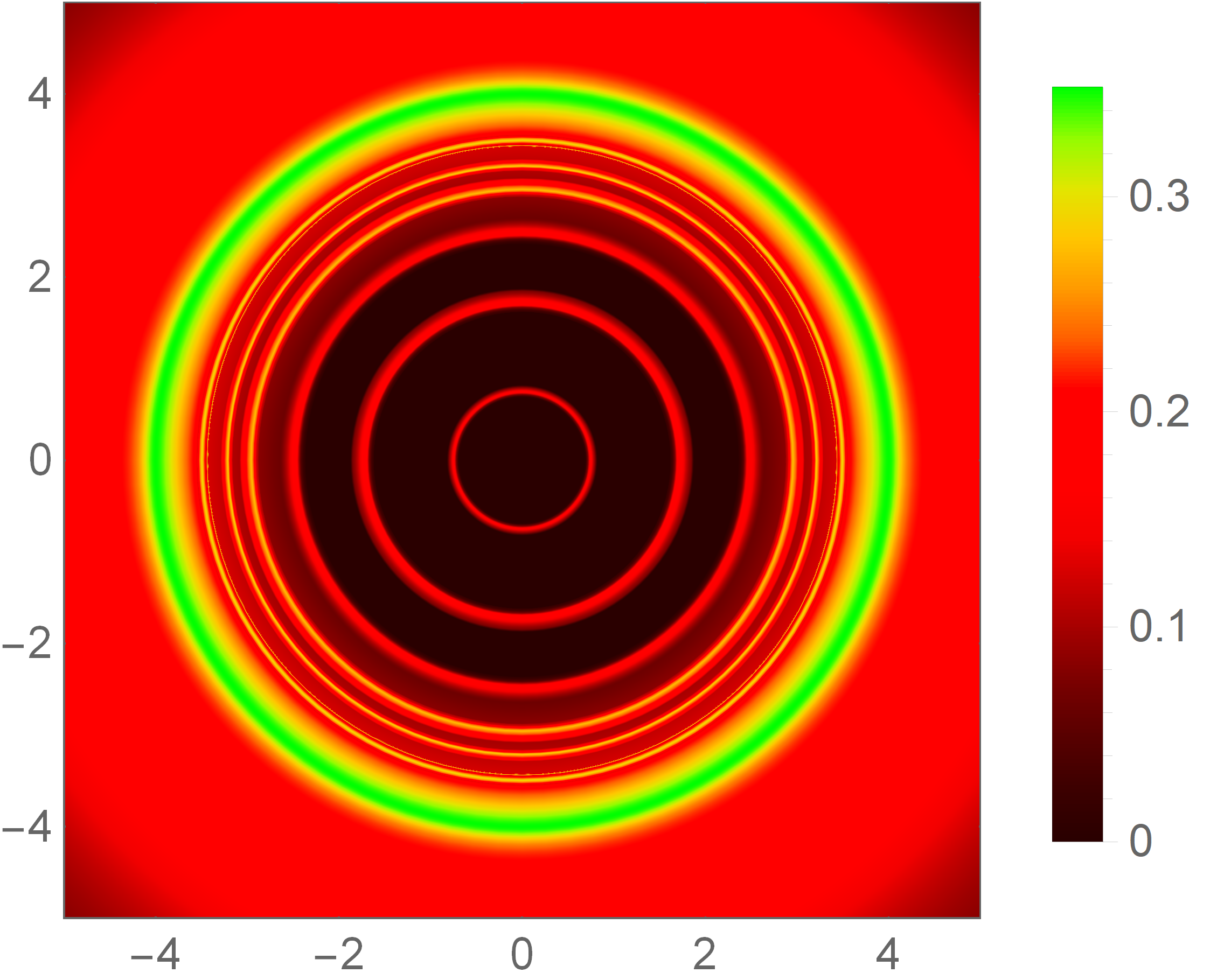}
\includegraphics[width=5.9cm,height=5.0cm]{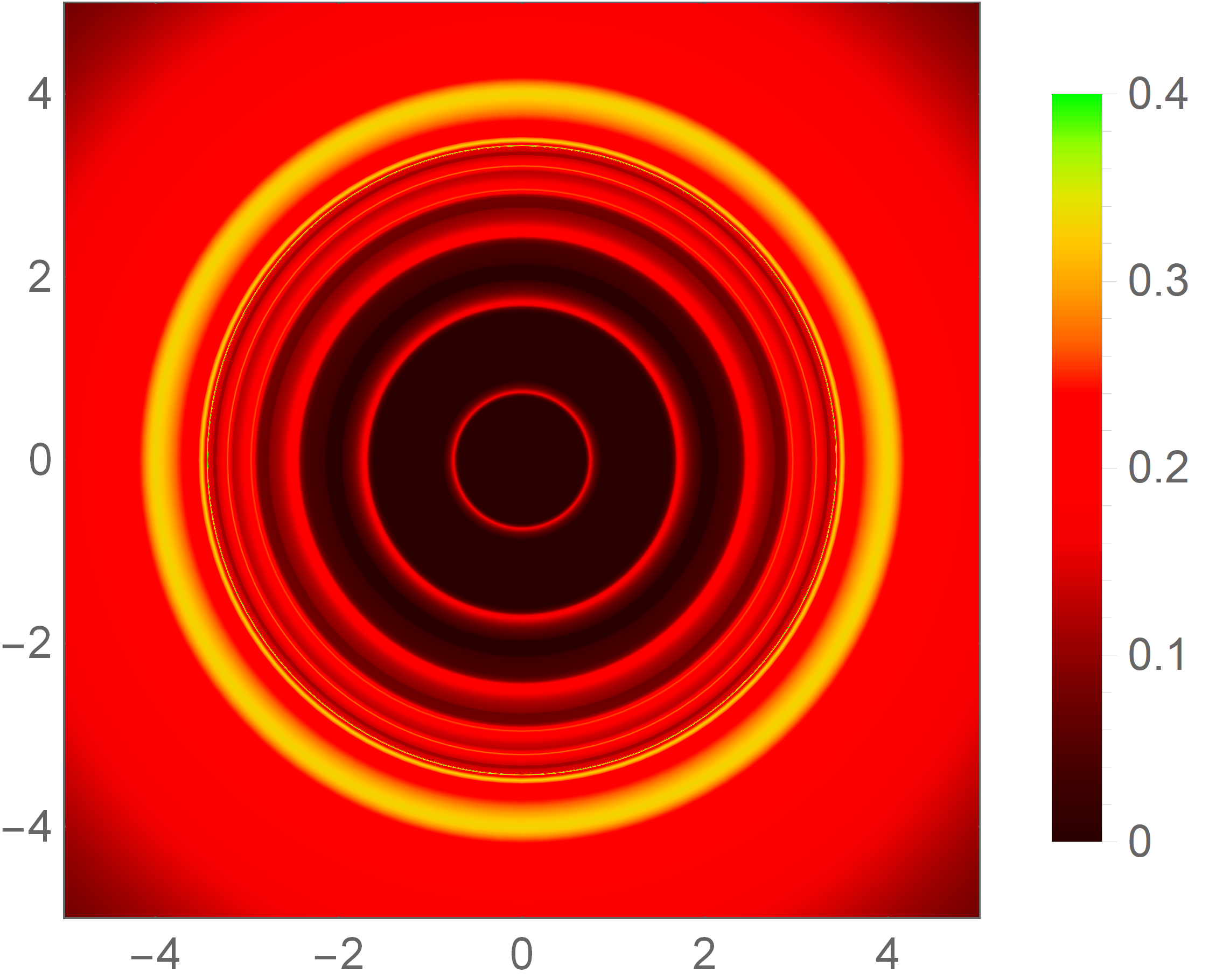}
\includegraphics[width=5.9cm,height=5.0cm]{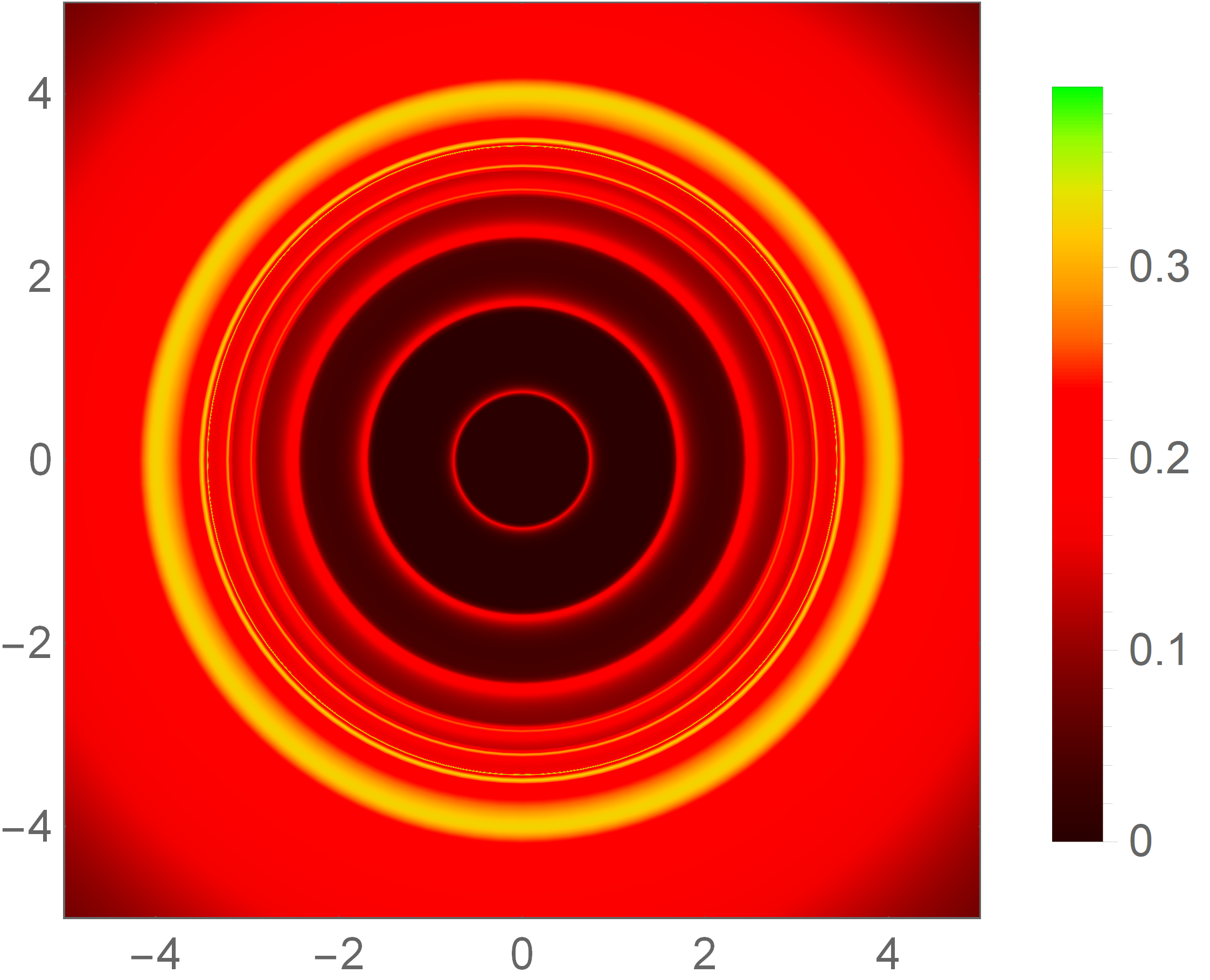}
\includegraphics[width=5.9cm,height=5.0cm]{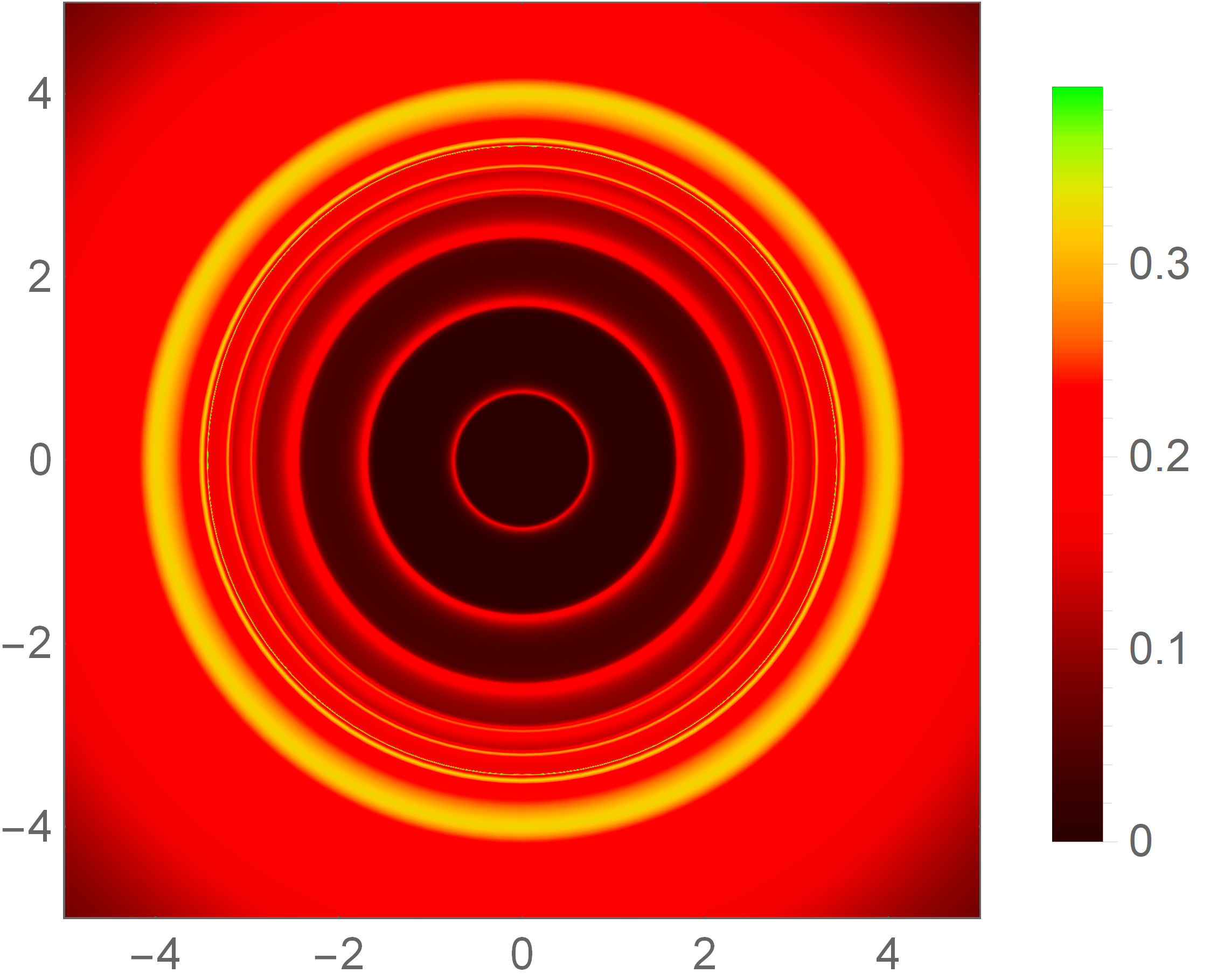}
\includegraphics[width=5.9cm,height=5.0cm]{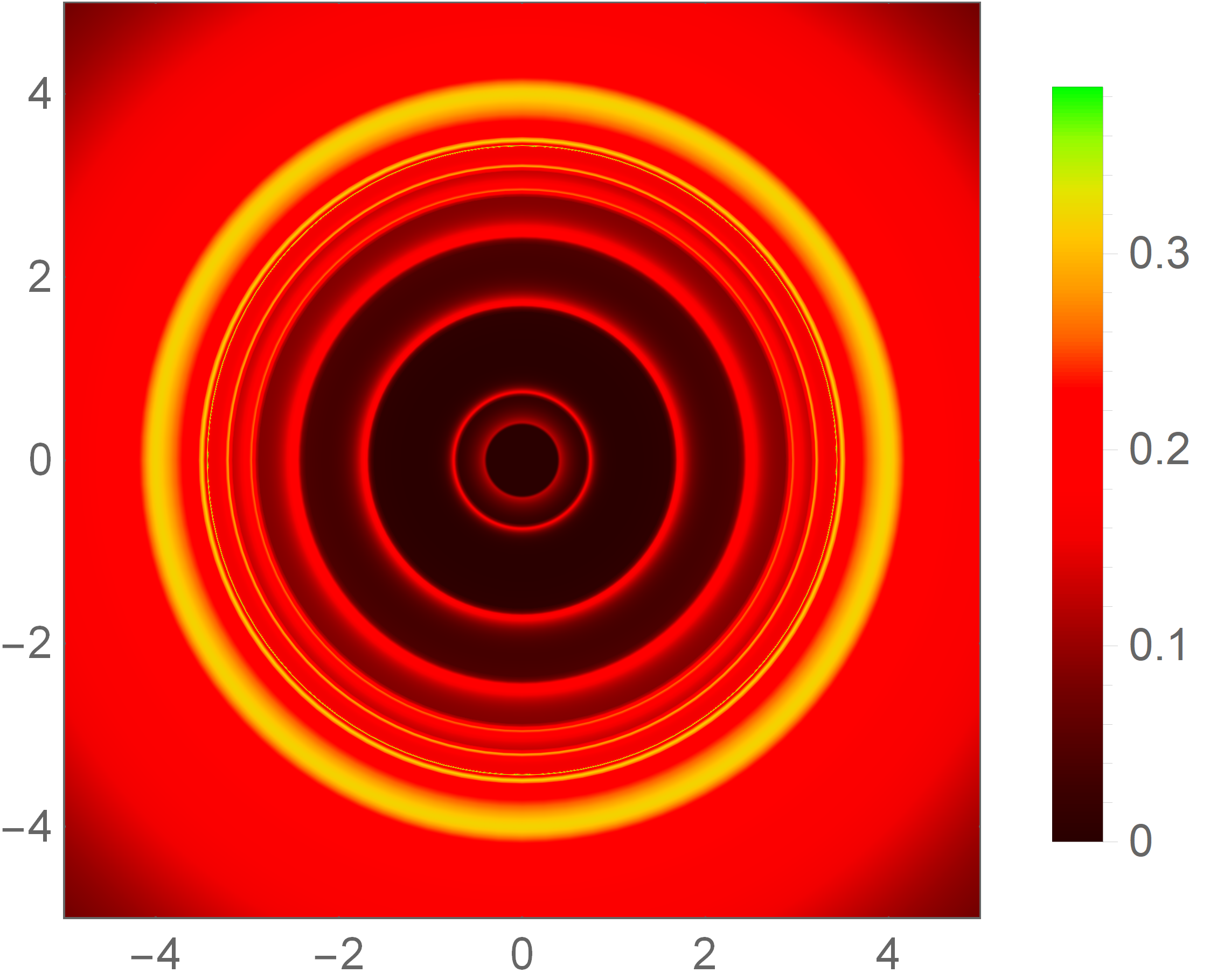}
\includegraphics[width=5.9cm,height=5.0cm]{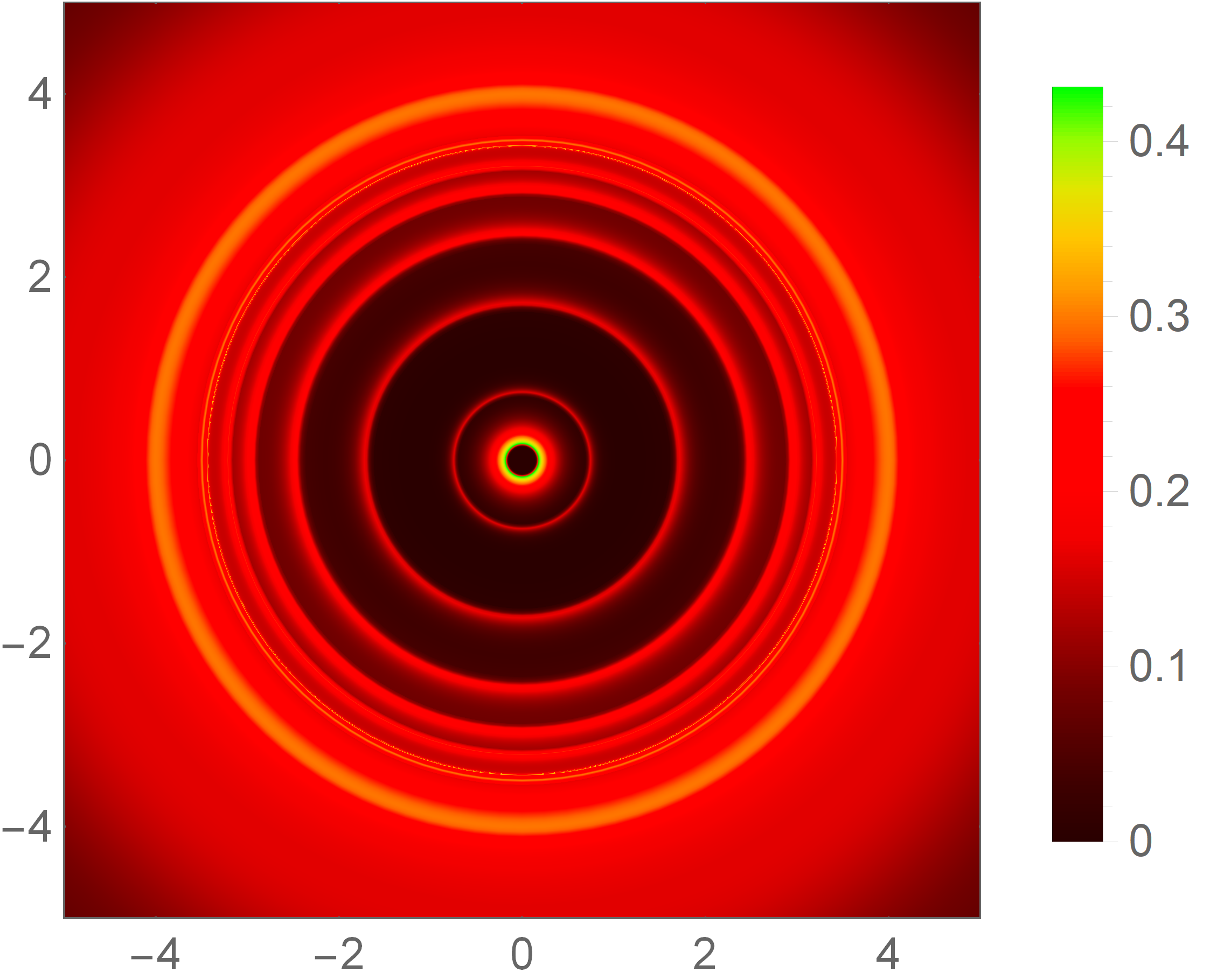}
\includegraphics[width=5.9cm,height=5.0cm]{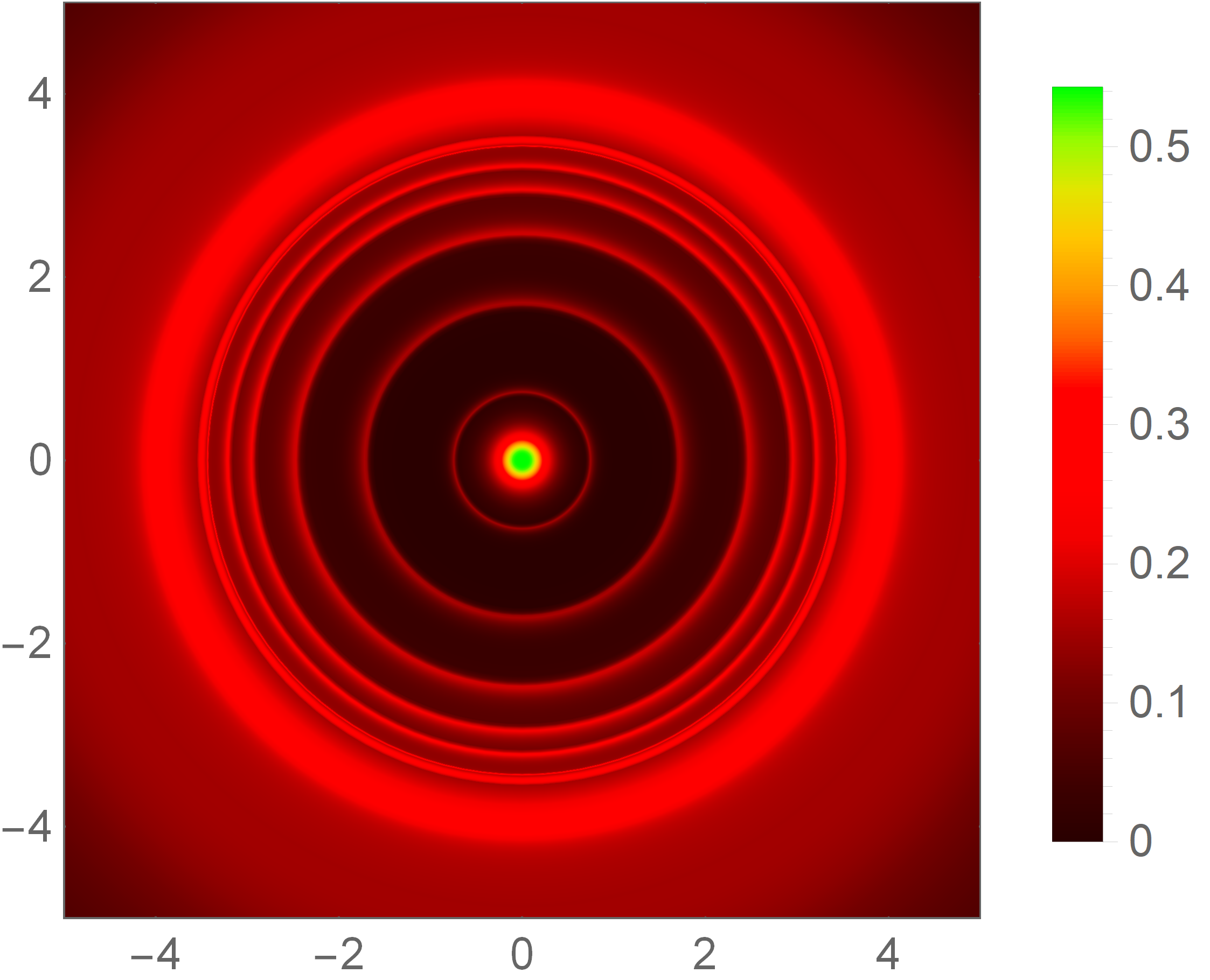}
\caption{The optical appearance of the SV object in the emission model (\ref{eq:I3}). Same notation as in Fig. \ref{fig:accmod1}.}
\label{fig:accmod3}
\end{figure*}

We next plug the above emission profiles into Eq.(\ref{eq:Iob}) to feed the light trajectories within our background geometries according to the data of the ray-tracing collected in the previous section. To get a first glimpse on the new features of the image, we first use the same example of Fig. \ref{fig:Iob} to depict a suitable cut of the associated multi-ring structure in Fig. \ref{fig:cut} (see Sec. \ref{eq:multiring} below for an interpretation and discussion of the observed structure). Next, we build the sequence of optical appearances of Figs. \ref{fig:accmod1}, \ref{fig:accmod2} and \ref{fig:accmod3}, corresponding to the three emission models (\ref{eq:I1}), (\ref{eq:I2}), and (\ref{eq:I3}),  respectively. In these figures we start the simulations by locating the inner edge of the disk, $r_{ie}$, at the ISCO radius, which is depicted in the top left figure of each of these plots. This choice of the initial surface is motivated, besides its obvious physical relevance, on the grounds that it maximizes the chances of neatly seeing at naked eye the multi-ring structure associated to the $2 \leq m \leq 6$ modes, as follows from Fig. \ref{fig:Iob}. Moving the value of $r_{ie}$ from the ISCO downward in discrete steps (left to right and top to bottom in all these plots) to produce a total of nine figures per emission model, another relevant value corresponds to the location of the critical curve, $r_m \approx 1.5528$ (middle left), until arriving to the very center of the object, $r=0$ (bottom right). From these plots one can observe the great influence of the choices of both the location of the inner edge of the disk (by comparing the different panels within the same figure) and the decay of the emission profile (by comparing the different figures) in the distribution of luminosity among the different rings.

\subsection{Multi-ring structure} \label{eq:multiring}

In the Schwarzschild solution of GR, the sequence of higher-order photon rings is on a one-to-one correspondence with the modes they are produced from, converging to the critical curve (in spherical models \cite{Falcke:1999pj}) or to the inner shadow (in both thin \cite{Chael:2021rjo} and thick models \cite{Vincent:2022fwj}). The luminosity of each photon ring is exponentially suppressed so rings beyond $m=3$ can be safely dismissed in their contributions to the optical appearance of the black hole.

From the plots \ref{fig:accmod1}, \ref{fig:accmod2} and \ref{fig:accmod3} one can notice several new features in the optical appearance of this SV object as compared to the Schwarzschild black hole. A general comment looking at the different panels from the figures is that a given mode $m$ may have associated more than one photon ring, and moreover they can be overlapped with each other. The net effect is that, in the images of the three emission profiles above, a maximum of up to eight rings (i.e. two of them are the usual ones associated to the lower-order modes, while the other six are genuine higher-order rings) can be separately seen, though this is only possible when the starting surface of the emission (given by the inner edge of the disk) is pushed to sufficiently far away distances (as in the example of Fig. \ref{fig:cut}, or in the more physical one of the $r_{ie}=r_{ISCO}$ in the top left panel in all figures below), with a sub-dominant influence on the choice of the decay of the emission (i.e., on the three models above). As $r_{ie}$ gets closer to the center of the object, the direct emission can (and actually does) superimpose  with one or more of such rings\footnote{Note that one can always split these contributions by viewing the distribution of the observed intensity rather than the full image, since in the former one clearly differentiates the peaks associated to every ring, see Fig. \ref{fig:Iob} for an example of this. Furthermore, adding more modes to the ray-tracing would result in additional peaks appearing at random locations due to the gravitational redshift, but their contributions to the total luminosity of the object can be utterly neglected due to their very reduced height.}  to yield fewer but wider rings. Moreover, while their location is almost unchanging irrespective of the location of $r_{ie}$, their luminosity can be significantly boosted depending on the degree of superimposition with the direct emission. Indeed, the choice of the emission profile plays the role of redistributing the luminosity of the direct contribution over those of the lower and higher-order rings in such a way that  the smoother the decay, the greater effect on this redistribution. One can easily appreciate it when going from Fig. \ref{fig:accmod1} (cubic decay of Eq.(\ref{eq:I1})) to Fig. \ref{fig:accmod2} (quadratic decay of model (\ref{eq:I2})), where the luminosity of the lower and higher-order rings of the latter is much more pronounced than in the former due to the larger spread of the luminosity of the direct contribution. This effect is strongly exaggerated in the model (\ref{eq:I3}) depicted in Fig. \ref{fig:accmod3}, where the larger spread of the emission profile infuses all the higher-order rings with additional luminosity at all choices of $r_{ie}$.

In view of this analysis, the shadow, understood as the dark central area of the image from which significantly much less luminosity comes out, is bounded by the innermost of the higher-order rings as long as the location of the inner edge of the disk does not penetrate deep enough into the region inside the critical curve (top and medium figures). When this is the case (bottom figures), the shadow will be bounded instead by the (gravitationally redshifted) location of the inner edge of the disk, at which the emission profile is truncated. Thus, below some value of $r_{ie}$, as it gets closer to the center of the object, $r=0$, the shadow's size gets diminished. In the limit $r \to 0$ one would arrive to a {\it shadowless} naked core (bottom right figure in all these plots) formed by those light rays that have acquired the divergent lens effect besides the little deflection at low enough values of $b$, though one could however argue that the infinite potential barrier of the effective potential (for both null and time-like trajectories) would actually prevent the particles making up the disk from penetrating down near $r=0$. Nonetheless, if we keep pressing upon this possibility, we find that the objects lying at the bottom of these plots contain a central luminous core with a tiny shadow inside and surrounded by a series of diluted photon rings, whose visibility depends strongly on the emission profile, from the almost non-visible rings of Figs. \ref{fig:accmod1} and \ref{fig:accmod2} to the net sequence of Fig. \ref{fig:accmod3}. The very last object of the sequence (i.e. when $r_{in} \to 0$) actually looks like a tiny luminous shadowless ball governing its petty multi-ring kingdom of photon rings.

Let us stress that the main ingredient behind the generation of higher-order ring images is the presence of both a photon sphere and an infinite potential at the center in the chosen solutions, together with the lack of an event horizon, which we recall it corresponds to the range $0.73576 \lesssim l<0.8$. Should we have chosen the branch of solutions below the lower limit a horizon would be present, and one would find the usual set of photon rings of the Schwarzschild solution (though modified in their locations and luminosities) converging to its inner shadow, while above the upper limit the photon sphere disappears and one find just the infinite potential deflecting all light rays  having no photon rings.

\section{Conclusion}
\label{conclusions}

In this work we have studied the images generated by a proposal to extend the Kerr black hole by a new type of (analytically tractable) family of compact objects, and which in the non-rotating limit degenerates into a modification of the effective potential of the Schwarzschild case by introducing an infinite slope at the center driven by a single new parameter. 

This simple toy-model merits in removing the curvature singularity at its center in terms of an asymptotically Minkowski core, and can act as a proxy for the presence of qualitatively new structures on shadow observations when accompanied  by the presence of a critical curve (i.e., a maximum on its effective potential), an anti-photon sphere, and the lack of an event horizon. In such a case, this SV object introduces new sets of trajectories of light rays turning more than one half-orbits around the compact object. When the main source of illumination is supplied by an optically thin accretion disk surrounding it, such light trajectories contribute to the image of the object by boosting the luminosity of the usual lower-order rings of the Schwarzschild solution (corresponding to those turning at most three half-orbits around the object) and creating a set of additional higher-order rings (associated to the light trajectories turning more than three times) whose contribution to the luminosity of the object cannot be longer  dismissed. This conclusion was reached by calling upon three geometrically thin analytical disk models in which the emission is truncated at the inner edge of the disk, where it actually takes its maximum value, smoothly decreasing outwards according to different decays. Within this setting, we found up to six higher-order rings (besides the two lower-order ones of the Schwarzschild solution) superimposed on top of the direct emission main ring, the size and luminosity of each being strongly influenced by the choice of the background geometry parameter, by the location of the inner edge of the disk, and by its assumed decay with distance.

The results found in this paper are in agreement with recent views in the field regarding the difficulty to disentangle the contributions from the background geometry and the astrophysics of the disk via the direct emission alone when the disk is assumed to be geometrically thin, while arguing in favour of the better opportunities present in the secondary rings, namely, the sequence of lower-order and higher-order rings. Indeed, the sharpness of higher-order ring images grants them their dominance in the high-frequency domain of the interferometric signal, and because of this fact, they could constitute smoking guns for the existence of new gravitational physics to be sought within future projects of very-long baseline interferometry, see e.g. \cite{Johnson:2019ljv,Gralla:2020srx,Carballo-Rubio:2022bgh} for a discussion on this point. Arguably, the model considered in this work is too contrived and modifies quantitatively far too much the basic features of the Schwarzschild images to represent a viable alternative to canonical black hole candidates, particularly under the light of recent  observations like those of the EHT (which actually infer a geometrically thick rather than thin accretion disk), but it supports the dire need to build a sort of {\it shadowgraphy}, namely, a thorough characterization of the qualitative possible shapes of the effective potential created by different background geometries. While this should be certainly possible for spherically symmetric space-times, the extensions of this programme to more realistic rotating space-times would face additional difficulties, since in such a case the integrability of the geodesic equations cannot be taken for granted, and similar radial and angular potentials as in the photon shell of Kerr solution do not necessarily exist in every possible rotating solution. Should one manage to build theoretically well supported alternatives to the Kerr picture that, in particular, might agree with the inferred features of the shadow caster of the M87 and Sgr A* observations while qualitatively departing from the shape of the Schwarzschild potential, our results point out that the presence of multi-ring images hidden in the main direct-emission ring could act as discriminators between both classes of objects.

\section*{Acknowledgements}

MG is funded by the predoctoral contract 2018-T1/TIC-10431. DRG is funded by the {\it Atracci\'on de Talento Investigador} programme of the Comunidad de Madrid (Spain) No. 2018-T1/TIC-10431. DS-CG is funded by the University of Valladolid (Spain), Ref. POSTDOC UVA20.  This work is supported by the Spanish Grants FIS2017-84440-C2-1-P, PID2019-108485GB-I00, PID2020-116567GB-C21 and PID2020-117301GA-I00 funded by MCIN/AEI/10.13039/501100011033 (``ERDF A way of making Europe" and ``PGC Generaci\'on de Conocimiento"), the project PROMETEO/2020/079 (Generalitat Valenciana), the project H2020-MSCA-RISE-2017 Grant FunFiCO- 777740, the project i-COOPB20462 (CSIC),  the FCT projects No. PTDC/FIS-PAR/31938/2017 and PTDC/FIS-OUT/29048/2017, and the Edital 006/2018 PRONEX (FAPESQ-PB/CNPQ, Brazil, Grant 0015/2019). This article is based upon work from COST Action CA18108, supported by COST (European Cooperation in Science and Technology). All images of this paper were obtained with our own codes implemented within Mathematica\circledR.

\end{document}